\pdfoutput=1 %

\documentclass[sigconf, nonacm]{acmart}

\newcommand\vldbdoi{XX.XX/XXX.XX}
\newcommand\vldbpages{XXX-XXX}
\newcommand\vldbvolume{14}
\newcommand\vldbissue{1}
\newcommand\vldbyear{2020}
\newcommand\vldbauthors{\authors}
\newcommand\vldbtitle{\shorttitle} 

\newcommand{\eat}[1]{}
\newcommand{\Var}{\mathsf{Var}}
\newcommand{\E}{\mathsf{E}}
\newcommand{\expe}{e^{\epsilon}}
\newcommand{\expehalf}{e^{\epsilon/2}}
\newcommand{\supp}{\operatorname{supp}}

\newcommand{\edit}[1]{#1}

\newcommand{\para}[1]{\smallskip\noindent\textbf{#1.}}

\usepackage[justification=centering]{caption}
\usepackage{subcaption}
\usepackage{booktabs}
\usepackage{dirtytalk}
\usepackage{cases}
\usepackage{mathtools}
\DeclarePairedDelimiter{\ceil}{\lceil}{\rceil}

\begin{document}

\title{Frequency Estimation under Local Differential Privacy [Experiments, Analysis and Benchmarks]}

\author{Graham Cormode, Samuel Maddock, Carsten Maple}
\affiliation{
\institution{University of Warwick, UK}
}
\email{ {g.cormode,s.maddock,cm}@warwick.ac.uk}

\eat{\author{Ben Trovato}
\affiliation{%
  \institution{Institute for Clarity in Documentation}
  \streetaddress{P.O. Box 1212}
  \city{Dublin}
  \state{Ireland}
  \postcode{43017-6221}
}
\email{trovato@corporation.com}

\author{Lars Th{\o}rv{\"a}ld}
\orcid{0000-0002-1825-0097}
\affiliation{%
  \institution{The Th{\o}rv{\"a}ld Group}
  \streetaddress{1 Th{\o}rv{\"a}ld Circle}
  \city{Hekla}
  \country{Iceland}
}
\email{larst@affiliation.org}

\author{Valerie B\'eranger}
\orcid{0000-0001-5109-3700}
\affiliation{%
  \institution{Inria Paris-Rocquencourt}
  \city{Rocquencourt}
  \country{France}
}
\email{vb@rocquencourt.com}

\author{J\"org von \"Arbach}
\affiliation{%
  \institution{University of T\"ubingen}
  \city{T\"ubingen}
  \country{Germany}
}
\email{jaerbach@uni-tuebingen.edu}
\email{myprivate@email.com}
\email{second@affiliation.mail}

\author{Wang Xiu Ying}
\author{Zhe Zuo}
\affiliation{%
  \institution{East China Normal University}
  \city{Shanghai}
  \country{China}
}
\email{firstname.lastname@ecnu.edu.cn}

\author{Donald Fauntleroy Duck}
\affiliation{%
  \institution{Scientific Writing Academy}
  \city{Duckburg}
  \country{Calisota}
}
\affiliation{%
  \institution{Donald's Second Affiliation}
  \city{City}
  \country{country}
}
\email{donald@swa.edu}
}
\begin{abstract}
Private collection of statistics from a large distributed population is an important problem, and has led to large scale deployments from several leading technology companies. 
The dominant approach requires each user to randomly perturb their input, leading to guarantees in the local differential privacy model. 
In this paper, we place the various approaches that have been suggested into a common framework, and perform an extensive series of experiments to understand the tradeoffs between different implementation choices. 
Our conclusion is that for the core problems of frequency estimation and heavy hitter identification, careful choice of algorithms can lead to very effective solutions that scale to millions of users. 
\end{abstract}

\maketitle

\begingroup\small\noindent\raggedright\textbf{PVLDB Reference Format:}\\
\vldbauthors. \vldbtitle. PVLDB, \vldbvolume(\vldbissue): \vldbpages, \vldbyear.\\
\href{https://doi.org/\vldbdoi}{doi:\vldbdoi}
\endgroup
\begingroup
\renewcommand\thefootnote{}\footnote{\noindent
This work is licensed under the Creative Commons BY-NC-ND 4.0 International License. Visit \url{https://creativecommons.org/licenses/by-nc-nd/4.0/} to view a copy of this license. For any use beyond those covered by this license, obtain permission by emailing \href{mailto:info@vldb.org}{info@vldb.org}. Copyright is held by the owner/author(s). Publication rights licensed to the VLDB Endowment. \\
\raggedright Proceedings of the VLDB Endowment, Vol. \vldbvolume, No. \vldbissue\ %
ISSN 2150-8097. \\
\href{https://doi.org/\vldbdoi}{doi:\vldbdoi} \\
}\addtocounter{footnote}{-1}\endgroup

\section{Introduction}

Collecting frequency statistics underpins a wide range of modern data analytics tasks, from simple popularity charts to more complex machine learning tools. 
If we can centrally collect and aggregate the raw data, then producing such statistics involves relatively simple data manipulation. 
However, in many cases the inputs can be considered sensitive, describing an individual's actions, preferences or characteristics.  
As such, the data subjects may be reluctant to divulge their true information, even if the overall statistics would be valuable and socially useful.  
For example, consider a mobile app that has access to a user's location.  
Understanding the overall population distribution, where people like to visit at particular times, and how they move around, has shown to be very valuable for social studies, urban planning, and tracking disease spread. 
However, any individual's location can be sensitive, allowing inferences to be made on their health (if they attend a clinic), religious preference (if they attend a place of worship), sexuality (if they attend a group or club), and more. 
For these reasons, users of the app would understandably be unwilling to share their location information freely with the app. 

In recent years, a number of privacy-preserving solutions to this frequency estimation question have been developed and deployed by large technology companies, such as Apple, Google and Microsoft. 
These deployments differ in the details, but share some common DNA; they are all based on the notion of Local Differential Privacy, an instantiation of the statistical model of differential privacy which is applied by each user directly to their own data.  
At the heart of these protocols, each user is asked a simple question about their data by a data analyst.  
The user then determines randomly whether to answer the question truthfully, or to pick some false response, according to a specified probability distribution.  
This randomization gives each user ``plausible deniability'', and protects their privacy.  
To fully specify a protocol, we need to determine what questions are asked, and what probability distribution is used over the set of possible answers.  
We also need to show how the data analyst can combine the answers to build a picture of the overall frequency distribution, and understand the accuracy of the results. 

In this overview and evaluation paper, we seek to draw together the various approaches (both theoretical and applied) that have been proposed for this core problem, and place them in a common framework. 
We show that protocols can be understood as combining four ``layers'': 
(1) A basic ``frequency oracle'' that allows frequencies to be estimated over a moderate sized discrete set of possibilities;
(2) A ``sketch'' that can reduce the domain size of possibilities;
(3) A ``heavy hitters'' method that can find frequent items within a large domain; 
(4) Post-processing techniques that enforce some constraints on the results to improve accuracy. 
Existing protocols can then be expressed by specifying a particular choice of algorithm for each of the four layers.  

Having described the choices in the four layers, we perform experiments that demonstrate the different behaviors of choices for each component, in terms of their accuracy and scalability (speed) as a function of privacy level, domain size, and other parameters. 
This leads us to recommend particular combinations for this task, and to deprecate others. 
Although not our main objective, we suggest some alternative choices that have not previously been considered (based on sampling hash functions), and show that these equal best-in-class performance. 
Meanwhile, we show that the first approaches that were introduced for these problems and popularized them are dominated by more advanced techniques that came afterwards. 

\paragraph{Outline.}
We proceed as follows: 
Section~\ref{sec:prelims} introduces the problem definitions and key technical concepts.
We describe the four layers of a private heavy hitters protocol in turn: 
first we summarize the ``frequency oracles'' which are the basis of all solutions in Section~\ref{sec:fo}. 
The other three layers of a scheme are domain size reduction (Section~\ref{sec:domain}), 
heavy hitter search (Section~\ref{sec:hh}), and post-processing (Section~\ref{sec:post}). 
We structure our experimental study into two parts: calibration and study of the basic frequency estimation task (Section~\ref{sec:freqexpts}), 
and comparative evaluation on the heavy hitters task (Section~\ref{sec:hhexpts}). 
We summarize our findings in Section~\ref{sec:concs}.

\section{Preliminaries}
\label{sec:prelims}
\subsection{Local Differential Privacy (LDP)}

The definition of differential privacy for a randomized procedure requires that the output distribution for inputs that are similar (``neighboring''). 
The additional requirement to satisfy local differential privacy is that this should apply to the output distribution of each data owner, rather than just on the overall output of a centralized procedure. 
More formally, we consider a collection of $n$ data owners, who each holds some input $x_i$. 
We assume that each item $x_i \in \mathcal{D}$ is drawn from some domain $\mathcal{D}$ of size $|\mathcal{D}|=d$.
We write $e^{(j)}$ as the basis vector of dimension $d$ such that 
$e^{(j)}_j=1$, and 0 otherwise. 
Each user will apply a local randomizer $\mathcal{R}$ to their input to obtain an output from a domain $\mathcal{Y}$, so that
$\mathcal{R} : \mathcal{D} \rightarrow \mathcal{Y}$.

\begin{definition}[Local differential privacy]
The local randomizer $\mathcal{R}$ is $\epsilon$-locally differentially private ($\epsilon$-LDP) if for all pairs of inputs $x\in\mathcal{D}, x'\in \mathcal{D}$
and outputs $y \in \mathcal{Y}$
we have that $\frac{\Pr[\mathcal{R}(x)=y]}{\Pr[\mathcal{R}(x')=y]} \leq \expe$. 
\end{definition}

The goal is to obtain the best accuracy for a particular task while guaranteeing  $\epsilon$-LDP. 
Further background on local differential privacy is presented in recent surveys and tutorials~\cite{survey1,survey2,tutorial}.

\subsection{Frequency Estimation and Heavy Hitters}

With the same setup as the previous section, we define the (raw) frequency of item $x \in\mathcal{D}$ to be
$f(x) = |\{ i : x_i = x \}|$. 
That is, we simply count the number of occurrences of item $x$ among the population. 
Natural variations consider a weighted version of this definition, but our focus here is on the unweighted (unit weight) notion. 
Our objective will be to obtain a (private, randomized) estimator $\hat{f}$ for $f$. 
We will typically be interested in estimators $\hat{f}$ that are \emph{unbiased}, that is $\E[\hat{f}(x)] = f(x)$, and which have small \emph{error} $|\hat{f}(x) - f(x)|$. 
When the estimator is unbiased, the mean squared error is equivalent to the variance, $\E[(\hat{f}(x) - f(x))^2]$. 
The \textit{heavy hitters} are items $x$ whose frequency is large --  this is often formalized to be those items which exceed a threshold of the total weight or total sum of squares, or are in the top-$k$ largest weights. 

\subsection{Hadamard Transformation}
\label{sec:hadamard}

Several private mechanisms depend on the Hadamard transformation, which is a particular instance of a discrete Fourier transform. 
It is an orthogonal transform $\phi$, described by a $d \times d$ matrix with $\phi_{i,j} = d^{-1/2} (-1)^{\langle i, j \rangle}$, where $d$ is a power of 2, and $\langle i, j \rangle$ counts the number of indices in the binary representations of $i$ and $j$ where both have 1 bit. 
Given a vector $x$ of dimension $d$, the Hadamard transformation of $x$ is the vector of coefficients
$\Theta = \phi x^T$.
For general $x$ this can be computed efficiently in time $O(d \log d)$, but we often apply this to sparse inputs and find one coefficient, which can be 
done with a constant number of 
mathematical
operations. 

\section{Frequency Oracles}
\label{sec:fo}
A Frequency Oracle (FO) is an LDP mechanism for frequency estimation. 
Oracles vary in their construction, accuracy guarantees, and the size of domain for which they are best suited. 
\edit{Our study builds upon \cite{wangetal} by comparing to additional oracles, and extending to cover the interaction of frequency oracles with domain size reduction (Section \ref{domain reduction}) and heavy hitter finding (Section \ref{sec:hh})}

\subsection{``Pure'' Protocols}

A useful abstraction introduced by Wang \textit{et al.}~\cite{wangetal} is the notion of a
`pure' protocol for frequency estimation. 
A protocol is considered to be pure if its output can be interpreted as expressing ``support'' for a subset of possible input values, encoded as $\supp(y)$. 
Pure protocols are characterized by two probabilities, $p^*$ and $q^*$, so that
$\Pr[\mathcal{R}(x) \in \{y : x \in \supp(y)\}] = p^*$,
which encodes the probability that input $x$ is mapped to an output that supports $x$;
and
$\Pr[\mathcal{R}(x') \in \{y : x \in \supp(y)\}] = q^* \text{ for all $x'\neq x$}$,
which encodes the (lower) probability that an input $x'$ is mapped to an input that supports $x$ for every $x'\neq x$. 

Many of the frequency oracles below are instances of a pure protocol, which makes their analysis convenient. 
The $\epsilon$-LDP condition is satisfied provided $p^*/q^* \leq \expe$. 
To aggregate the responses from a pure protocol to make an estimate $\hat{f}(x)$, for each report $y_i = \mathcal{R}(x_i)$, we compute 
$\hat{f}_i(x) = \frac{\mathcal{I}_{x\in \supp(y_i)} - q^*}{p^*-q^*}$
where $\mathcal{I}_{P}$ is the function that is 1 if $P$ is true, and 0 otherwise. 
We can verify that 
\[ 
\E[\hat{f}_i(x)] = {(\E[\mathcal{R}(x_i)\in \{y: x \in \supp(y)\}] - q^*)}/{(p^* - q^*)}
\]
which simplifies to 1 if $x_i=x$, and 0 if $x_i \neq x$.

The contribution to the (conditional) variance of this estimator for each user $i$ can be computed for the two cases, as: 
\begin{numcases} 
{\Var[\hat{f}_i(x)] = 
\frac{\Var[\mathcal{I}_{x\in \supp(y_i)}]}{(p^* - q^*)^2}=}
  \frac{p^*(1-p^*)}{(p^* - q^*)^2}\!\!\!\!\!\!\!\!&\text{if} $x=x_i$~~\label{eq:pvar}\\
  \frac{q^*(1-q^*)}{(p^* - q^*)^2}\!\!\!\!\!\!\!\!&\text{if} $x\neq\!x_i$~~\label{eq:qvar}
\end{numcases}
The overall estimate $\hat{f}(x)$ is the sum of the individual estimates, and its variance is the sum of these variances. 
The exact value of the overall estimate's variance will depend on the true frequency of $x$ in the input. 
If we assume that the domain is large and no input dominates the others, then this expression is determined by the expression in $q$, and we can focus on the contribution to the variance from the $x\neq x_i$ case in \eqref{eq:qvar}.

\subsection{Direct Encoding (DE)}
\label{sec:de}
The idea of direct encoding (also known as randomized response) is to set the output range to be the same as the input, i.e., $\mathcal{Y} = \mathcal{D}$,
and to have some probability of providing a ``false'' response, while maximizing the likelihood of giving a ``truthful'' response~\cite{rappor,wangetal}. 
If we set the truthful probability to be
$\Pr[\mathcal{R}(x) = x]=p$, 
then for any $y\neq x$ we must have 
$\Pr[\mathcal{R}(x) = y]=(1-p)/(d-1)=q$
and LDP requires $p/((1-p)/(d-1)) \leq \expe$. 
Equating and rearranging this, we obtain
$p = \frac{\expe}{\expe + d - 1}$ and $q = \frac{1}{\expe + d -1}$. 
This yields a pure protocol with $p^* = p$ and $q^*=q$.
To build an unbiased estimator for this mechanism, the contribution from each report is
$\hat{f}_i(x) = \frac{(\expe+d-1)\mathbb{I}_x(\mathcal{R}(x_i)) - 1}{\expe-1}$, 
where $\mathbb{I}_x(\cdot)$
is the indicator function for whether its argument is equal to $x$. 
It then follows that
$\E[\hat{f}_i(x)] = \mathbb{I}_x(\mathcal{R}(x_i))$, and that
\[
\Var[\hat{f}_i(x)] = (\mathbb{I}_x(\mathcal{R}(x_i))p(1-p) + (1-\mathbb{I}_x(\mathcal{R}(x_i))q(1-q))
\left(\frac{\expe+d-1}{\expe-1}\right)^2.\]
\noindent
The variance term corresponding to case $x_i\neq x$ from \eqref{eq:qvar} for DE is
$\Var[\hat{f}(x)]=\frac{q(1-q)}{(p-q)^2} = \frac{\expe + d -2}{(\expe-1)^2}.$

\subsection{Unary Encodings (OUE, SUE)}
\label{sec:oue}
Unary encoding methods interpret the user's input $x_i$ as a one-hot $d$-dimensional vector $e^{(x_i)}$, and independently perturb each bit. 
Hence, $\mathcal{Y} = \{0,1\}^d$. 
Suppose we retain the (sole) input 1 with probability $p$, and flip each 0 to a 1 with probability $q$. 
This provides a pure protocol with $p=p^*$ and $q=q^*$, where we have the freedom to choose $p$ and $q$ subject to satisfying 
$p(1-q)/(1-p)q \le \expe$ to ensure $\epsilon$-LDP\footnote{Considering neighboring inputs which differ in the location of the solitary 1 bit.}. 
``Symmetric'' unary encoding (SUE), as used in RAPPOR~\cite{rappor} sets $p+q=1$, yielding $p=\frac{\expehalf}{\expehalf + 1}$ and $q=\frac{1}{\expehalf+1}$.
The corresponding variance is
$pq/(p-q)^2 = \expehalf/(\expehalf-1)^2$.
``Optimized'' unary encoding (OUE, \cite{wangetal}) aims to minimize \eqref{eq:qvar} subject to 
$p(1-q)/(1-p)q = \expe$.
Rearranging this sets $p=q \frac{\expe}{1+q\expe - q}$, and hence
$p-q = \frac{(\expe-1)q(1-q)}{1+q\expe - q}$ and 
so \eqref{eq:qvar} gives the variance contribution as
$\frac{(1 + q(\expe-1))^2}{(\expe-1)^2q(1-q)}$. 
We can minimize the factor in $q$ by taking the partial derivative with respect to $q$ and equating to zero. 
This leads us to choose $q=1/(\expe+1)$, and the expression for
\eqref{eq:qvar} simplifies to $\frac{4\expe}{(\expe-1)^2}$.
This fixes $p=\frac12$, i.e., the encoding retains the sole original input 1 with constant probability, while flipping zeros only with smaller probability $1/(\expe+1)$. 

\subsection{Local Hashing (BLH, OLH, FLH)}
As $d$ gets larger, the dependence on $d$ for DE is undesirable, and subsequent approaches remove this dependency.
The local hashing approach has each user pick a hash function with which to encode their input, and send back the hash function along with the hashed input encoded via direct encoding~\cite{wangetal}. 
Here, the domain of the hash is chosen to be $g$, with $g$ chosen as a function of $\epsilon$, typically much smaller than $d$. 
That is, user $i$ picks a (universal) hash function $h_i$ that maps 
$[d] \rightarrow [g]$. 
Next, the user applies direct encoding to the hashed value $h_i(x_i)$, and sends this output along with the description of $h_i$.
From this, the aggregator can obtain an unbiased estimator for $\hat{f}(h_i(x))$. 
This gives a pure protocol, where the probability that $x_i$ is mapped to the hash value that supports $x_i$ is 
$p^* = \frac{\expe}{\expe + g -1}$, 
while the probability that $x\neq x_i$ is mapped to that hash value becomes the uniform chance $q^* = 1/g$. 

The variance contribution due to \eqref{eq:qvar} 
$\frac{(\expe + g - 1)^2}{(g-1)(\expe-1)^2}$.
A simple case is Binary Local Hashing (BLH), which just sets $g=2$, and obtains variance
$\frac{(\expe+1)^2}{(\expe-1)^2}$.
However, by choosing 
$g-1  = \expe$ (Optimal Local Hashing, OLH) we minimize the factor
$\frac{(\expe + (g-1))^2}{(g-1)}$, and hence obtain a variance of
$\frac{4\expe}{(\expe-1)^2}$. 
Note that this leads us to choose 
$p^*=\frac12$, i.e., we report the true hash value with probability half. 
A limitation of the OLH approach is that it can be very slow. 
This is since we need to make $O(nd)$ hash function calls to calculate which domain elements a user's perturbed item contributes frequency towards. 

Here, we propose a ``fast'' heuristic modification which aims to speed up the procedure, dubbed Fast Local Hashing (FLH).
On the client-side, instead of sampling a hash function uniformly at random from some universal hash family, we introduce a new parameter $k^\prime$ and restrict clients to uniformly choosing from $k^\prime$ hash functions. 
Hence, we sacrifice some theoretical guarantees on accuracy in order to achieve computational gains on the server-side aggregation. 
In practice, it turns out that even for small values of $k^\prime$, FLH performs reasonably well.

The protocol is largely unchanged from OLH. 
    On the client-side, we sample a hash function uniformly at random from $\{h_1, \dots h_{k^\prime}\}$.
    On the server-side, we  pre-compute a $k^\prime \times d$ matrix where each row corresponds to a hash function and each column refers to a domain value. 
    We set the $(i,j)^{th}$ entry to $h_i(j)$ to reduce the total number of hash function calls from $O(nd)$ to $O(k^\prime d)$.
    The server collates all reports for the same hash function, and process them in a batch. 
To gain a speedup over OLH, we must choose $k^\prime \ll n$. 
In our experiments (Section~\ref{sec:expts:flh}), we empirically validate FLH, and show that it approaches OLH in accuracy when only sampling a small number of hash functions, achieving a substantial speed up.

\subsection{Hadamard Encodings (HM, HR)}
\label{HM}
We next consider two related approaches based on the Hadamard transform. 
The Hadamard mechanism (HM) samples one (or more) coefficients from the Hadamard transform of the users input (treated as a sparse vector), and applies direct encoding 
to the result~\cite{appledp,practicalhh}. 
So $\mathcal{Y} = \{-1, +1\}$ and
given a user's input $x_i \in \mathcal{D} = [d]$, we sample an index $j \in [d]$, and compute the (scaled-up) coefficient
$\theta_j^{(i)} = \phi_{x_i,j} = (-1)^{\langle x_i, j\rangle}$ (see Section~\ref{sec:hadamard}). 
Then with probability $p = \frac{\expe}{1+\expe}$, the mechanism reports $(j, \theta_j^{(i)})$, otherwise it reports $(j, -\theta_j^{(i)})$. 
Since $\frac{p}{1-p} = \frac{\expe}{1+\expe} \cdot \frac{1+\expe}{1} = \expe$, this meets the LDP guarantee. 

An advantage of HM
is that it is much faster to process reports from users, since the (fast) Hadamard inverse transform can be applied to the (aggregated) reports. 
To build an unbiased estimator for frequencies, we take the contribution of the inverse of the unbiased estimator of the reported $\hat{\theta}_j = \sum_{i} \theta_j^{(i)}$. 
The unbiased estimator for $\theta_j$ is ${\hat{\theta}_j}/{(2p-1)}$.
For a given $x$, to estimate $f(x)$, we sum the contribution to $f(x)$ from all reports. 
That is, 
$\hat{f}_i(x) = \sum_j \phi_{x,j} \cdot {\hat{\theta}_j^{(i)}}/{(2p-1)}$
and $\hat{f}(x) = \sum_i \hat{f}_i(x)$ as before.
We can interpret the $j$'th row of the Hadamard matrix as defining a (universal) hash function on the input domain with range $g=2$. 
The analysis for BLH applies, and
the variance bound \eqref{eq:qvar} is
$\frac{(\expe+1)^2}{(\expe-1)^2}$. 

To improve this bound when $\expe$ is large, we can 
sample $t$ Hadamard coefficients, to produce a hash function with $g=2^t$ possible outcomes.  
This preserves the result with probability $p^* = \frac{\expe}{\expe+2^t-1}$, and otherwise perturbs it uniformly. 
The resulting variance from \eqref{eq:qvar} is 
$\frac{(\expe + 2^t - 1)^2}{(2^t-1)(\expe-1)^2}$. 
If we can choose $t$ so that $2^t - 1 = \expe$,\footnote{This is possible when $\expe$ is moderately large, i.e., $\expe=3, 7, 15. \ldots$} then we obtain the optimal variance of $\frac{4\expe}{(\expe-1)^2}$. 
We can aggregate quickly by interpreting the hash value as $t$ Hadamard coefficients, each of which is 
transmitted correctly with probability
$\frac{\expe}{\expe + 2^t-1} + \frac{2^{t-1}-1}{\expe + 2^t -1} \cdot \frac{2^{t-1}-1}{2^t -1} = \frac{\expe + 2^{t-1}-1}{\expe + 2^t-1}$. 
For $\expe=3$ and $t=2$, this fraction is $2/3$.

\eat{\textit{
In expectation over the random choice based on $p$, we have
\begin{align*}
\E_p[\hat{f}(x)] & = \frac{\phi_{x,j}}{2p-1}\E_p[\phi_{x_i,j}]\\
&= \frac{\phi_{x,j}}{2p-1} (p \phi_{x_i,j} + (1-p)(1-\phi_{x_i,j}))\\
&= \phi_{x,j} \phi_{x_i,j}
\end{align*}
We can observe that if $x=x_i$, then 
$\phi_{x_i,j}\phi_{x,j} = \phi^2_{x,j} = 1$ for all $j$, whereas
if $x \neq x_i$, then over the choice of $j$ we have that
$\Pr[\phi_{x_i,j}\phi_{x,j}=1] = \Pr[\phi_{x_i,j}\phi_{x,j}=-1] = \frac12$. 
Hence, we can write that $\E[\hat{f}(x)] = e^{x_i}_x$, as required.

For the variance, we find that for $x\neq x_i$, the mean squared error is $(\frac{1}{2p-1})^2$, whereas for $x=x_i$, it is 
$\frac{4p(1-p)}{(2p-1^2)}$. 
Since we have $p=\frac{\expe}{1+\expe}$, we have
$1/(2p-1)^2 = (\frac{\expe+1}{\expe-1})^2$, and 
$\frac{4p(1-p)}{2p-1}^2=\frac{4\expe}{(\expe-1)^2}$
}
}

\label{sec:hr}
The Hadamard response protocol, HR,~\cite{hadamardresponse} also makes use of the Hadamard matrix, and is very similar to HM.
Each user with input $x_i$ 
will report the index of some Hadamard coefficient, i.e., some $j$, where the corresponding coefficient value is 
$\theta_j = \phi (e^{(x_i)})^T$.  
Equivalently, the set of all Hadamard coefficient values $\Theta$ correspond to the $x_i$'th column of Hadamard matrix $\phi$. 
With probability $p = \frac{\expe}{\expe+1}$, we report a $j$ such that $\theta_j=+1$, otherwise (with probability $q=\frac{1}{\expe+1}$), 
we report a coefficient $j$ with $\theta_j = -1$. 
This again represents a pure protocol, since by properties of the Hadamard transform, $p^* = p$, while $q^* = \frac12$. 
Hence, we compute our variance bound as $(\frac{\expe+1}{\expe-1})^2$.
For higher privacy (smaller epsilon), a variant protocol 
is based on the Hadamard matrix, but is no longer pure, since the probability of $x'$ being mapped to the support of $x$ is no longer the same for all $x'$.  
This construction achieves a variance bound of $O(\frac{\expe}{(\expe-1)^2})$.

\subsection{Summary of Frequency Oracles}

\begin{table}[t]
    \caption{Comparison of Frequency Oracle costs}
    \centering
    \begin{tabular}{l|lll}
         FO &  Variance bound & Decode time & Communication \\
         \hline
         DE & $O((\expe + d-2)/(\expe-1)^2)$  & $O(n +d)$ & $O(\log d)$\\
         SUE & $O(\expehalf/(\expehalf-1)^2)$ & $O(nd)$ & $O(d)$\\
         OUE & $O(\expe/(\expe-1)^2)$ & $O(nd)$ & $O(d)$\\
         BLH & $O((\expe+1)^2/(\expe-1)^2)$ & $O(nd)$ & $O(\log d)$\\
         OLH & $O(\expe/(\expe-1)^2)$ & $O(nd)$ & $O(\log d)$\\
         FLH & (heuristic) & $O(k'd)$ & $O(\log d)$\\
         HM & $O((\expe+1)^2/(\expe-1)^2)$ & $O(n+d)$ & $O(\log d)$\\
         HR & $O(\expe/(\expe-1)^2)$ & $O(n+d)$ & $O(\log d)$
    \end{tabular}
    \label{tab:fo}
\end{table}

The concept of a frequency oracle is an abstraction of a private protocol allowing us to accurately estimate the frequency of items from a domain. 
The listed approaches achieve different combinations of accuracy, speed and communication costs in order to guarantee $\epsilon$-LDP. 
We summarize the key properties in Table~\ref{tab:fo}. 

Across all frequency oracles, the variance obtained is quite similar, with a dependence on $\frac{1}{(\expe-1)^2}$. 
When $\epsilon$ is suitably small, we have
$\expe \approx 1 + \epsilon$, and so ${(\expe-1)^{-2}} \approx {\epsilon^{-2}}$. 
The numerator of the variance term is typically either $(\expe+1)^2$ (for basic Hadamard mechanism), or $4\expe$ (for Hadamard response, optimized local hashing and optimized unary encoding). 
Since $(x+1)^2 - 4x = (x-1)^2 > 0$ for $x>1$, 
so $4\expe < (\expe+1)^2$, and hence we would expect these last methods to obtain better accuracy, as we test in Section~\ref{sec:freqexpts}.

\section{Domain size reduction}
\label{sec:domain}
\label{domain reduction}
When working with data over large domain sizes $d$, performing a search for frequent items can be slow, and incur more false positives as we query more possibilities.  
A natural approach is to encode the input in a smaller domain, 
via techniques such as ``sketching''~\cite{CormodeYi20}. 
We can first encode each user's input into a small sketch, then apply the frequency oracle to each sketch, so that an aggregator can reconstruct a sketch to query. 
Since the oracles described above all rely on the user's input being sparse (i.e., encoded by a 1-hot vector),
we seek sketches which are sparsity-preserving. 
Each sketch is defined by an array with $r$ rows and $c$ columns, and some hash functions which map items to the array. 
Next, we survey three sketch approaches that have been used for this purpose. 

\para{Bloom Filter}
The Bloom filter (BF) is a well-known way to represent a set of items~\cite{Bloom:70}. 
It makes use of $k$ hash functions $h_1 \ldots h_k$ which map 
$[d] \rightarrow [m]$. 
This generates a one-dimensional binary array of length $m$, whose $j$'th entry is 1 if there is a hash function $h_\ell(x) = j$. 
Encoding a single user's item $x_i$ means that we obtain an encoding with at most $k$ 1's set. 
Frequency oracles can be applied by first sampling one of the hash functions $h_\ell$, and then processing the resulting one-hot encoding of $h_\ell(x_i)$. 
Decoding the result and obtaining an estimate from the noisily reconstructed Bloom filter can be done e.g., 
using regularized regression as a heuristic~\cite{rappor}.

\para{Count-Min Sketches}
\label{cm-sketch}
The Count-Min (CM) sketch is similar to the Bloom Filter, but now hash function $h_\ell$ maps to 
row $\ell$ of a $r \times c$ sized array~~\cite{countmin}. 
Item $x$ is then mapped to entry 
$[\ell, h_\ell(x)]$ 
for each $\ell$, so the sketch can be treated as $r$ one-hot vectors, one for each of the $r$ hash functions. 
The user first samples $\ell$ uniformly from $\{1 \ldots r\}$, and encodes $h_\ell(x_i)$ using a frequency oracle. 
The aggregator reconstructs a noisy sketch, and can estimate $f(x)$ by considering the weights associated with $h_\ell(x)$: this can be via taking the minimum, the mean, or the median (giving the Count-Mean or Count-Median sketch, respectively, with appropriate corrections to make the estimation unbiased~\cite{appledp,CormodeYi20}). 
The variance introduced due to sketching depends on $c$, the length of each row.  
Smaller $c$ and $r$ reduces the cost of storing and analyzing the sketch, but larger $c$ improves the accuracy. 

\para{Count Sketches}
The Count sketch (CS) follows the description of the CM sketch, but with one further twist: another hash function $g_\ell$ determines whether to represent $x$ with a $+1$ or $-1$ value~\cite{ccfc}.  
That is, the user samples $\ell$ uniformly from $r$, and encodes 
$h_\ell(x_i)$ with a weight of $g_\ell(x_i)$, where $g_\ell(x_i)=\{-1,+1\}$. 
This is most easily handled by the Hadamard mechanism, which handles negative inputs without any modification.  
Other FOs can be extended to handle negative weights, such as by handling positive and negative weights separately. 
To estimate $f(x)$, the aggregator takes the weight associated with $h_\ell(x)$, multiplied by $g_\ell(x)$. 
The variance due to sketching also depends on $1/c$, as in the previous case.

\section{Heavy Hitter Search}
\label{sec:hh}
Domain size reduction 
reduces the size of the objects manipulated in frequency estimation protocols. 
However, if the original domain within which the items reside is very large, it is still costly for the data analyst to search through the set of all possibilities to find those with high frequency.  
For example, the input items may be long strings (such as URLs), and it would be too slow to enumerate all strings up to a fixed length. 
This is known as the ``heavy hitters'' problem, and has been addressed extensively in the (non-private) data streaming setting. 
Approaches that work in the local privacy model are based on ideas from coding theory and group testing.

\para{Sequence Fragment Puzzle (SFP)}
The core idea of the ``sequence fragment puzzle'' (SFP) approach is that 
for a long string that is frequent within a population, every substring will also be (at least as) frequent~\cite{appledp}. 
Naively, we might try to find all frequent substrings and glue them back together.  
However, this risks creating false positives.
For example, suppose 
``apples'' and ``orange'' are both frequent
(6 character) strings. 
We could break these into 3 character substrings, as 
``app'', ``les'', ``ora'' and ``nge'', which will all be frequent. 
But when reconstructing, we should avoid reporting non-frequent strings, such as ``orales'' or ``ngeapp''. 
A heuristic approach 
is for each user to provide information on pairs of disjoint substrings that co-occur together in their string -- e.g., dividing strings into four substrings so that one user may reporting on their first and third substring, another on their third and fourth. 
This gives the aggregator partial information about which substrings co-occur frequently, allowing statistical inference to rebuild the whole strings~\cite{fanti16}. 

The SFP method gives a principled solution with two extra steps. 
The first is to provide the index of each substring within the string, and the second is to also tag each substring with a hash of the whole string. 
In our example, the pieces might be represented as
(1, ``app'', hash(apples)), (2, ``les'', hash(apples)), 
(1, ``ora'', hash(orange)), (2, ``nge'', hash(orange)). 
If an analyst detects these four tuples as frequent then, provided there are no hash collisions, 
the only way to combine them into strings yields the desired ``apples'' and ``orange''. 
In the LDP setting, each user takes their string and chooses a substring, then encodes this substring concatenated with the hash value via a frequency oracle (possibly with domain size reduction). 
The index of the substring not need be kept private, and can be sent without perturbation. 
This allows the analyst to find which substrings are globally frequent by enumerating all possibilities, and put these together to solve the `puzzle'. 

There are several parameters to choose for SFP:
(1) The length of the substrings to consider. 
Too short, and there is a lot of work to recombine the strings; too long, and the time cost of the analyst to search the frequency oracles is too high.  
Apple's implementation uses substrings of length 2-3 characters over the Roman alphabet.  
(2) The size of the hash function. 
This should be large enough to avoid collisions between frequent strings, but not so large that it makes the concatenated message large.  Apple's implementation uses an 8-bit hash.  
(3) How each user should select a substring to report. 
The suggested approach is to pad each user string to the same length, and uniformly sample a single index to report. 
In Apple's example, strings are 10 characters long, and we sample one of the five 2-character strings starting at an odd location. 

Reporting a single substring per user achieves better accuracy than reporting multiple substrings with a shared privacy budget. 
To avoid false positives, we can additionally maintain a frequency oracle over the full strings (at some privacy cost). 

\para{Hierarchical Search (PEM, TH)}
An alternative approach to finding heavy hitters is hierarchically based. 
Considering the inputs as strings, observe that if a string is a heavy hitter, then every prefix will be at least as frequent within the population. 
So we can build frequency oracles on prefixes of increasing length, and use these to step towards the full length strings. 
The key to making this efficient is the converse of the above statement: if a prefix is not a heavy hitter, then no extension of the prefix can be a heavy hitter either. 

The idea of hierarchical methods is to first enumerate all prefixes of some fixed length (say, 2 or 3 alphabet characters), and estimate their frequency with an appropriate frequency oracle. 
This yields a subset of heavy hitter prefixes. 
The next phase uses a frequency oracle on prefixes of a longer length, and estimates the frequency of all possible extensions of the heavy hitter prefixes. 
This is repeated until we reach the limiting length of prefixes for which a frequency oracle has been built. 
Variants on this idea are known under various names -- prefix extending method (PEM)~\cite{pem}, TreeHistogram (TH)~\cite{practicalhh}, and PrivTrie~\cite{privtrie}, but we group them together under the hierarchical search banner.  
The chief difference is how they report information: in PEM, each user is equally likely to report each prefix length, while in TH the privacy budget is split evenly to report on both the full string and 
and a uniformly chosen shorter prefix. 
As with SFP, there are several parameters to choose: 
(1) The initial prefix length, and how long to extend the prefix by in each phase; and 
(2) How each user should decide what information to report. 
A natural choice is for each user to uniformly select a prefix length from those specified, and only encode their input once.

\para{Error-correcting codes}
We briefly mention that other approaches have been proposed based on applying error-correcting codes (ECC) to allow recovering the heavy hitter strings.  
This general approach was first presented by Bassily and Smith~\cite{bassilysmith}, and used in the Bitstogram protocol~\cite{practicalhh} to give improved theoretical bounds. 
However, we omit this approach from further consideration for brevity, since it has not been widely implemented.

\section{Post-processing}
\label{sec:post}
\label{post-processing}
We finally consider how additional post-processing can be done on the outputs of frequency estimation and heavy hitter protocols with the aim of improving accuracy. 
Post-processing is useful, since the output of frequency oracles can be quite noisy: we can obtain negative frequencies, or outputs which sum to more the number of inputs.  
\edit{ In \cite{wangConsistency}, Wang \textit{et al.} present an experimental comparison of a large range of post-processing technique. 
We adopt a subset of the best of these techniques and consider as follows}:
\newline
\edit{(1) \textit{Base}}: No modification of the estimated frequencies.\newline
\edit{(2)   \textit{Base-Pos}}: We round all negative frequency estimates to $0$. \newline
\edit{(3)  \textit{Norm-Sub}}: We round negative estimates to $0$. For the rest of the values, we add/subtract some constant $\delta$ to ensure that $\sum_{v\in\mathcal{D}_{> 0}} (\tilde{f}(v) + \delta) = n$ where $\mathcal{D}_{>0} = \{v: \tilde{f}(v) > 0 \}$.\newline
\edit{(4)  \textit{Probability Simplex}}: Here we project our frequency oracle onto the probability simplex, ensuring that $\sum_{v\in\mathcal{D}} \tilde{f}(v) = n, \tilde{f}(v) \geq 0$. See \cite{HR} and \cite{prob-simplex} for more information.\newline
\edit{(5)  \textit{Base-Cut}}: When estimating the whole domain we sort our frequency estimates in decreasing order and keep them until we get a total frequency which is $>n$. At this point we round every remaining estimate down to $0$.

Note that applying some of these post-processing steps can be time consuming when the domain is large. 
Base-Pos is straightforward, but other post-processing requires enumerating all possible items in the domain to determine normalization parameters, and so
cannot be applied for frequency oracles over large input domains.

\begin{figure*}[t!]
\centering
  \subfloat[\label{fig:group1_eps} Varying $\epsilon$, fixing $d=1024, n=100,000$]{%
       \includegraphics[width=0.33\linewidth]{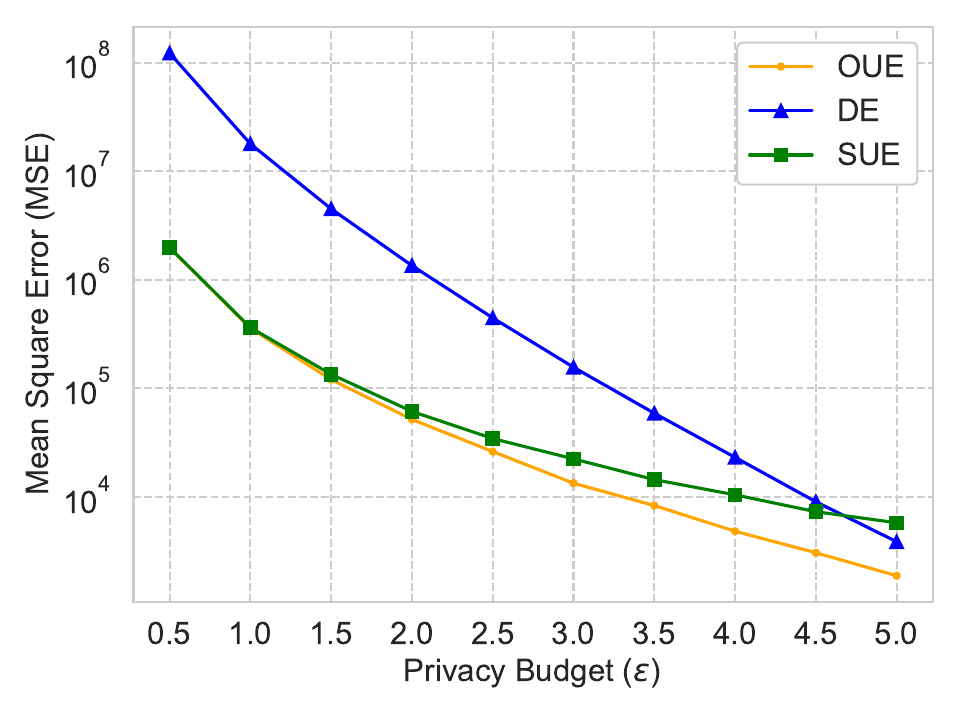}}
  \subfloat[\label{fig:group1_d} Varying $d$, fixing $\epsilon=3, n=100,000$]{%
        \includegraphics[width=0.33\linewidth]{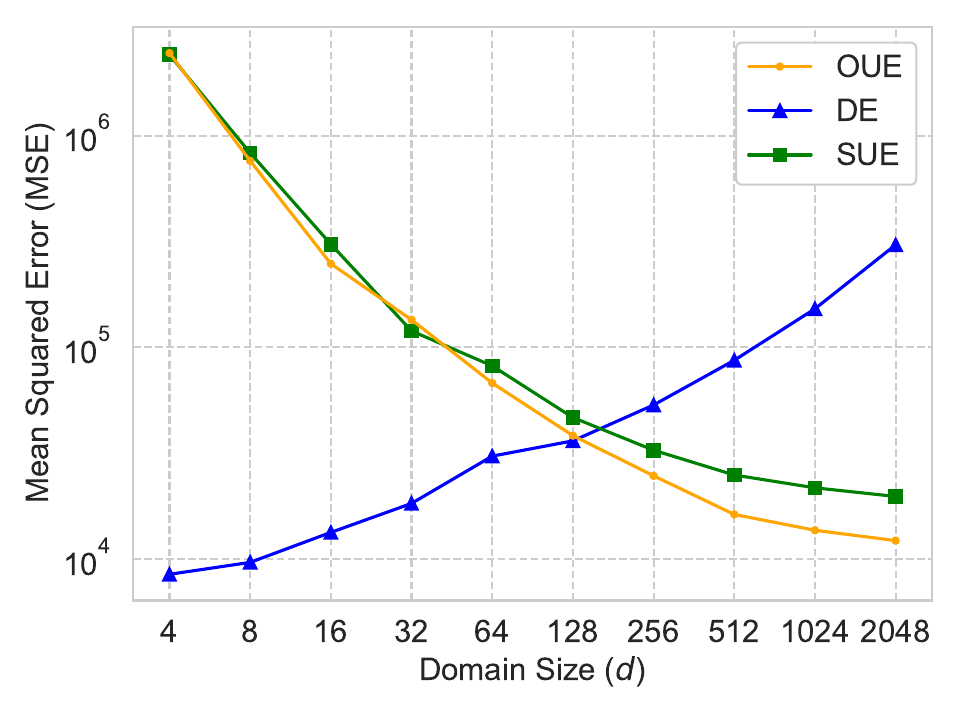}}
  \subfloat[\label{fig:group1_time} Total time taken as $d$ varies]{%
        \includegraphics[width=0.33\linewidth]{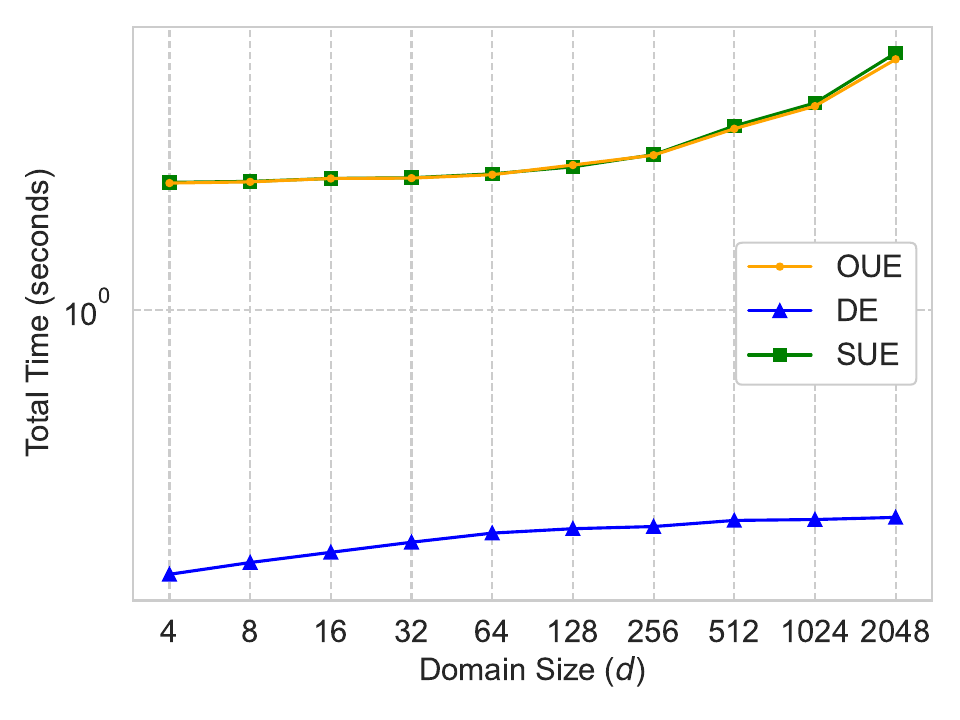}}
  \caption{DE, SUE, OUE Experiments}
  \label{fig:group1} 
\end{figure*}

\section{Frequency Estimation Experiments}
\label{sec:freqexpts}
Our first experiments focus on the ability of different frequency oracles and their variants to estimate frequencies. 
We are mainly interested in measuring the mean squared error (MSE) of the frequencies estimated over the whole domain i.e., $\frac{1}{d} \sum_{x \in \mathcal{D}} (\hat{f}(x) - f(x))^2$, which also serves as an empirical estimation of the variance. 

All experiments were implemented in Python 3.7.4 and run on a Windows PC, with an AMD Ryzen 5 3600 3.6GHz CPU and 16GB of RAM. The code is publicly available \footnote{\url{https://github.com/Samuel-Maddock/pure-LDP}}.
Unless stated otherwise all, experiments were run 5 times with their results averaged. 
We test the frequency oracles mainly on synthetic data generated from a Zipf distribution with skewness $s=1.1$ \footnote{
\edit{Additional experiments (not shown) used other input distributions, and did not change our conclusions, although more uniform inputs lead to lower accuracy. }}

\subsection{Direct and Unary Encoding Methods}
\label{sec:expts:fo:de}
We start by comparing Direct Encoding (DE), Symmetric Unary Encoding (SUE) and Optimal Unary Encoding (OUE) (Sections~\ref{sec:de} and \ref{sec:oue}). 
In Figure \ref{fig:group1_eps} we see the effects of varying the privacy budget $\epsilon \in \{0.5, 1, \dots, 5\}$ while fixing $d=1024, n=100{,}000$. 
Although the domain is not very large, OUE/SUE both  outperform DE due to the high amount of noise needed in DE. 
For relatively low values of $\epsilon$ (high privacy) both OUE and SUE perform quite similarly, however for larger values of $\epsilon$ we can see OUE has the lower MSE matching the theoretical results.

In Figure \ref{fig:group1_d} we vary the domain size $d \in \{2^2, 2^3, \dots, 2^{11}\}$ while fixing $\epsilon=3, n=100{,}000$. 
For very small domains, DE has much lower variance than OUE/SUE, however at around $d=2^6$ OUE/SUE start to outperform DE. 
This agrees with the theoretical results, since the variance of DE scales in $d$ as we require significantly more noise in order to maintain LDP across the domain.

In practice DE is fast and simple to implement, since all that needs to be returned by a client is a single value in the domain. 
OUE/SUE require perturbing  a $d$-length one-hot vector and sending it to the server. 
When the domain is large, this is impractical in terms of communication cost. Our results corroborate those of \cite{wangetal}.

\label{sec:expts:flh}
\begin{figure}[t]
  \centering
  \includegraphics[width=0.85\linewidth]{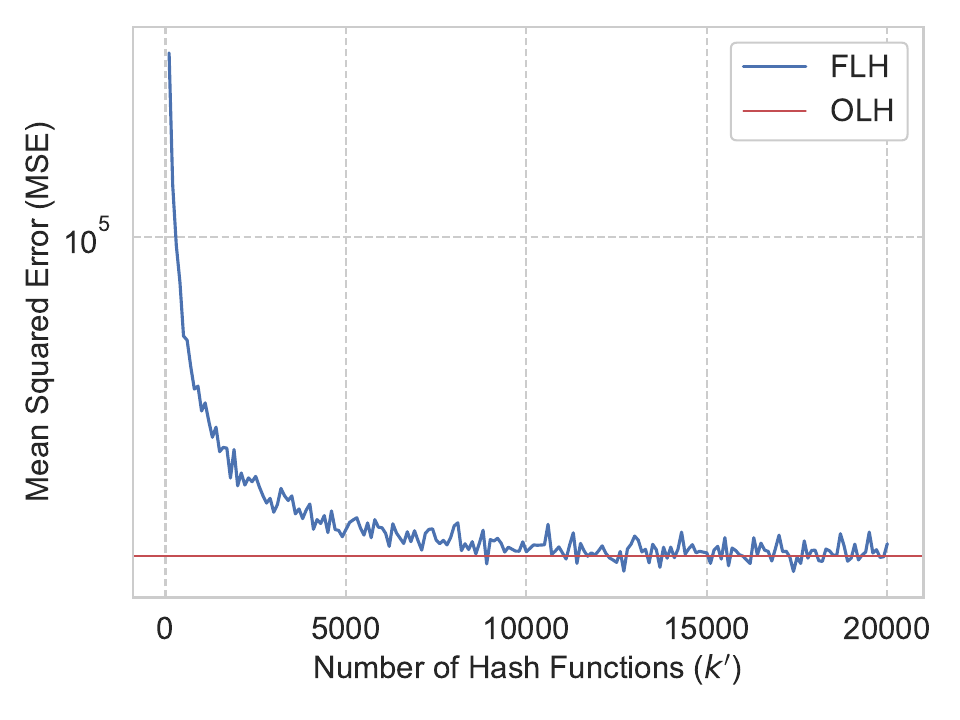}
  \caption{Effect of varying $k^\prime$ in FLH, $d=500, n=1,\!000,\!000, \epsilon=3$}
  \label{fig:group2_k}
\end{figure}

\begin{figure*}[t!]
\centering
  \subfloat[\label{fig:group2_eps} Varying $\epsilon$, fixing $d=1024, n=100,000$]{%
       \includegraphics[width=0.333\linewidth]{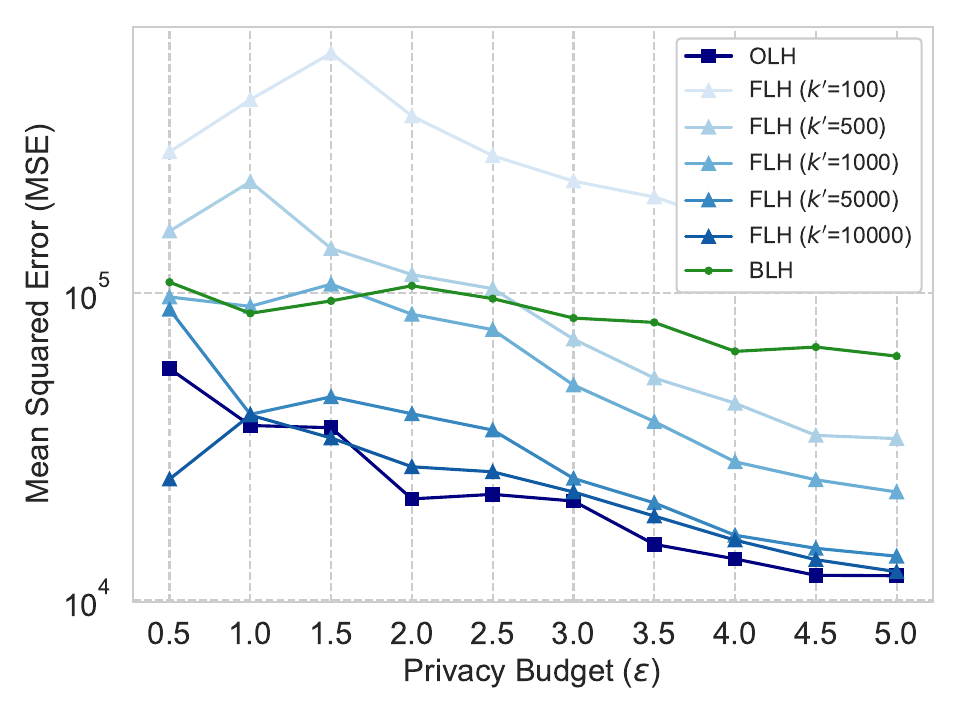}}
  \subfloat[\label{fig:group2_d} Varying $d$, fixing $\epsilon=3, n=100,000$]{%
        \includegraphics[width=0.333\linewidth]{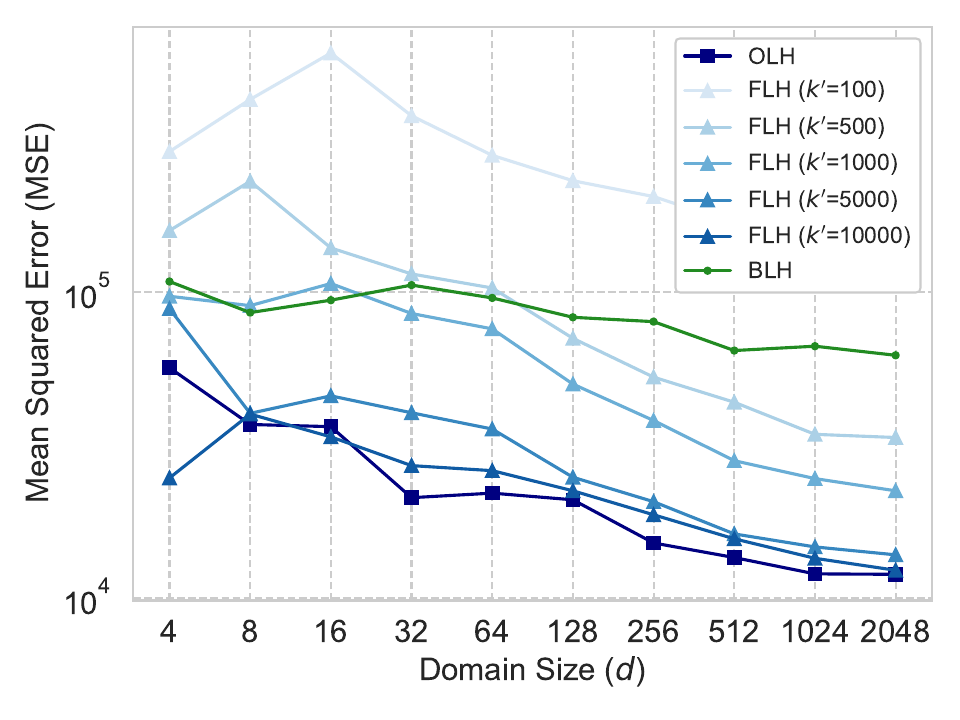}} 
  \subfloat[\label{fig:group2_time} Total time taken as $d$ varies]{%
    \includegraphics[width=0.333\linewidth]{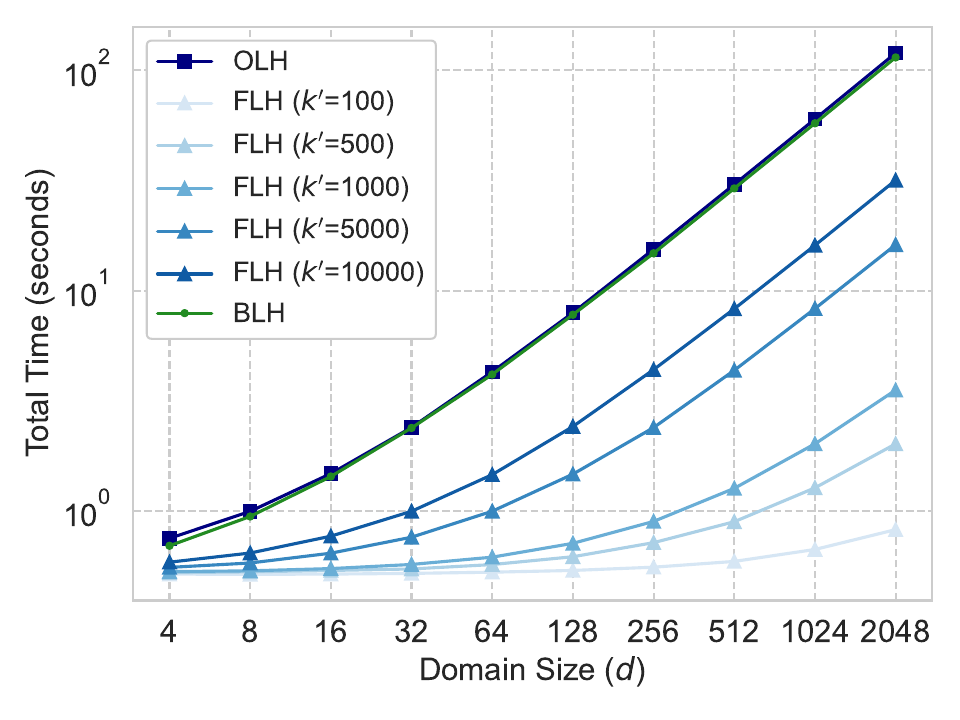}}
  \caption{Local Hashing (BLH, OLH, FLH) Experiments}
  \label{fig:group2} 
\end{figure*}

\subsection{Local Hashing Methods}
\label{sec:expts:fo:lh}
We compare Optimised Local Hashing (OLH) with Binary Local Hashing (BLH) (i.e., using a hash function with $g=2$ possible values) and with Fast Local Hashing (FLH). 
The parameter $k^\prime$ in FLH determines the number of hash functions used, with larger $k^\prime$ resulting in more accurate estimates at the expense of slower server-side aggregation.
In Figure \ref{fig:group2_k} we fix $d=500, n=1{,}000{,}000$ and vary $k^\prime$ between $100$ and $20{,}000$ in increments of $100$. 
The blue line represents the MSE of FLH as we vary $k^\prime$, and the red line represents the average OLH error on the same data. 
As $k^\prime$ increases, the MSE of FLH approaches that of OLH. 
At around $k^\prime=10{,}000$ the accuracy of FLH is indistinguishable from that of OLH. 
When $k^\prime=10{,}000$, FLH's aggregation is  $\approx 6$x faster than OLH.

Figure \ref{fig:group2} shows experiments for varying both $\epsilon$ and $d$ for OLH, BLH and FLH with $k^\prime= 100, 500, 1000, 5000, 10{,}000$. 
Figure~\ref{fig:group2_eps} shows the effect of varying $\epsilon \in \{0.5, 1, \dots, 5\}$ with $d=1024, n=100,000$. 
We observe that the MSE of BLH is consistent for different $\epsilon$ values, and as noted by \cite{wangetal} this is likely due to the use of a binary hash function losing a lot of information, making BLH relatively robust to low $\epsilon$ values. 
OLH has the lowest MSE followed closely by the FLH oracles in decreasing order of their $k^\prime$ value, as we would expect. 
We see similar relative behavior as we vary $d$ in Figure \ref{fig:group2_d}. 

Meanwhile, in Figure \ref{fig:group2_time}, we plot the total time taken across all client/server calculations as we vary $d$. 
Since we are enumerating the entire domain, this naturally has a linear dependency on $d$.
Nevertheless, we can see the effect that small values of $k$ in FLH have in terms of reducing the total time taken compared to OLH and BLH, which both have to iterate over $n$ different hash functions. 
A smaller $k^\prime$ parameter reduces the aggregation time of the server, but both client-side perturbation and server-side estimation take the same amount of time.

\begin{figure*}[t]
 \subfloat[\label{fig:group3_t} Varying $t$ and $\epsilon$ values for HM, fixing $d=1024,  n=100{,}000$]{%
        \includegraphics[width=0.333\linewidth]{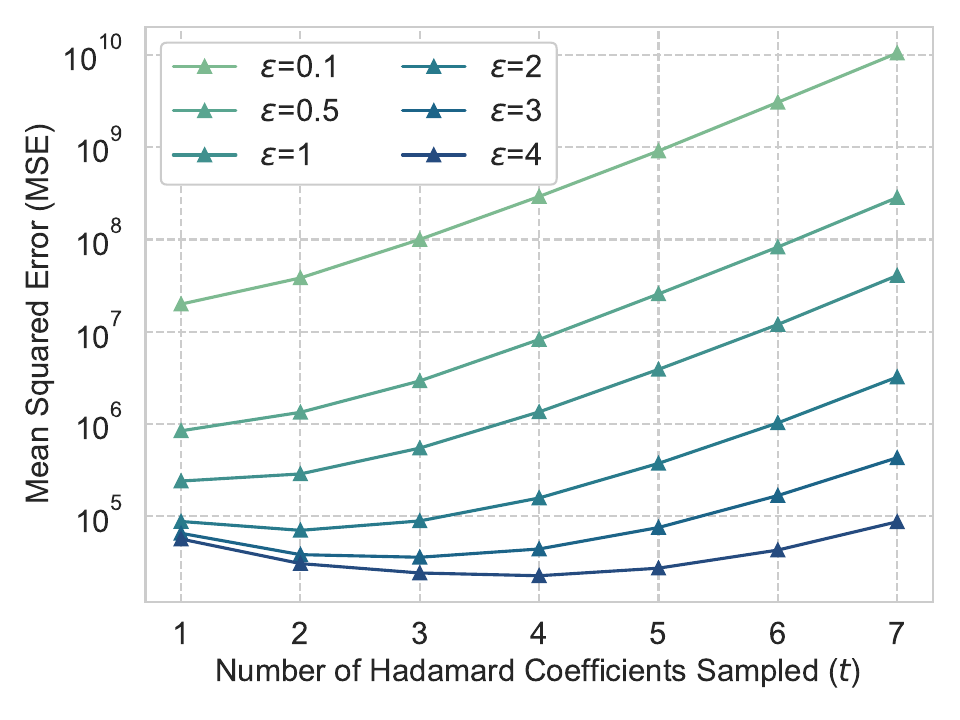}} 
  \subfloat[\label{fig:group3_eps} Varying $\epsilon$, fixing $d=1024, n=100{,}000$]{%
       \includegraphics[width=0.333\linewidth]{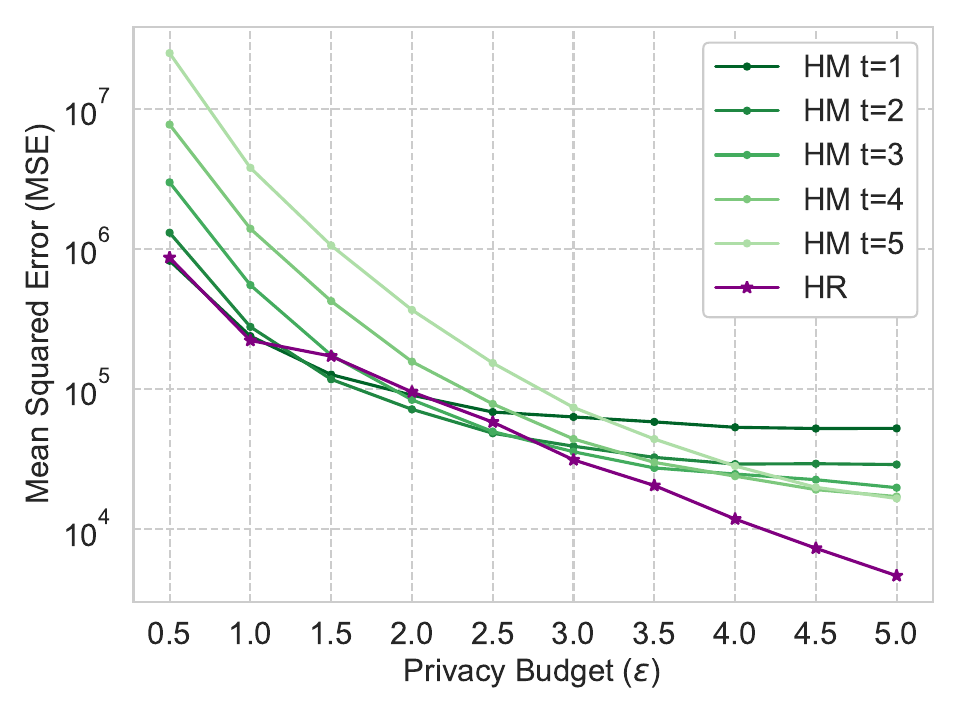}}
    \subfloat[\label{fig:group3_d} Varying $d$, fixing $\epsilon=3, n=100{,}000$]{%
        \includegraphics[width=0.333\linewidth]{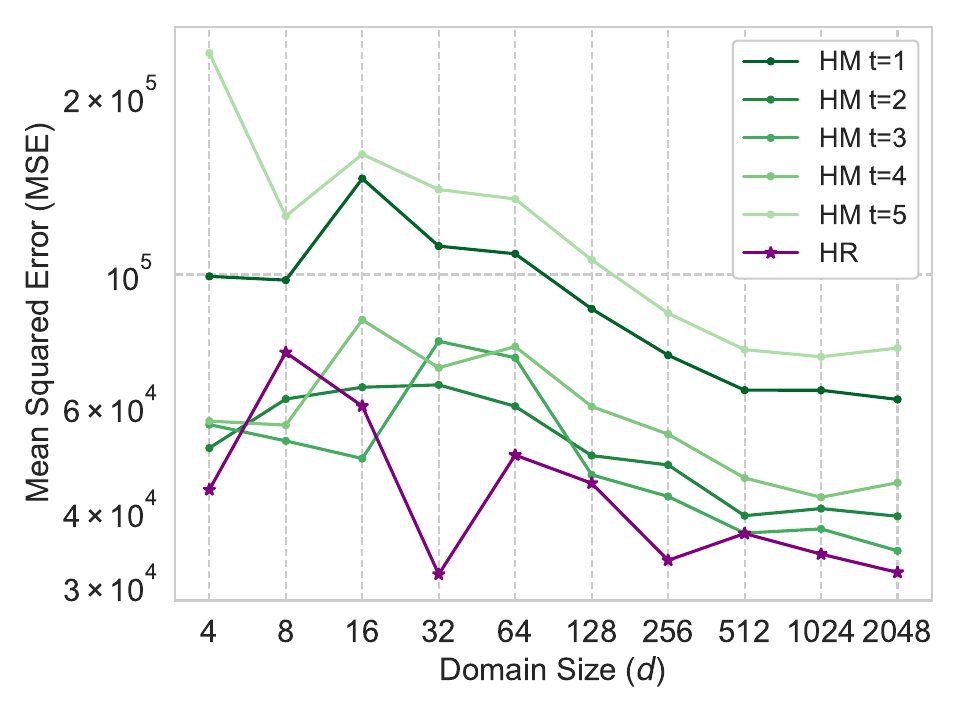}}\vspace*{-2mm}\caption{Hadamard Experiments}\label{fig:group3}%
\end{figure*}

\subsection{Hadamard Methods}
\label{sec:expts:fo:hm}
We compare the two Hadamard-based approaches (Section~\ref{HM}): Hadamard Mechanism (HM) and Hadamard Response (HR). 
The main parameter of the Hadamard Mechanism is the number of Hadamard coefficients to sample, which we denote as $t$. 
In Figure~\ref{fig:group3_t} we fix $d=1024, n=100{,}000$ while varying $t$ along the $x$-axis with each line representing a different value of $\epsilon$. 
We see that for low values of $\epsilon$ (high privacy) that the optimal value is $t=1$, whereas
for $\epsilon =3$, $t=3$ has the lowest MSE. Generally we see optimal values for $t=\ceil{\epsilon}$, consistent with our analysis of this mechanism. 

In Figure \ref{fig:group3_eps} we vary $\epsilon$ while fixing $d=1024, n=100{,}000$. 
We compare the MSE of HR with the HM with $t=1,2,3,4,5$. 
We can see that for the smaller $\epsilon$ values HR and HM perform very similarly when using the optimal values of $t=1,2$. 
Additional experiments were performed in the low privacy regime ($\epsilon < 1$) and we found that the performance of HR and HM were almost identical but we omit these plots for brevity. 
For higher $\epsilon$ values ($\epsilon>3$), HR starts to outperform the HM oracles, which is quite notable for methods which appear so similar on paper. 
In Figure \ref{fig:group3_d} we vary $d$ while fixing $\epsilon=3, n=100{,}000$. We can again see that HM $t=3$ has the best MSE for $\epsilon =3$, but still more than that of HR for the large domains. 
For very small domains, the behavior is quite variable, but nevertheless HM appears preferable in this regime.

\begin{figure*}[t]
  \subfloat[\label{fig:group4_eps} Varying $\epsilon$, fixing $d=1024, n=100{,}000$]{%
       \includegraphics[width=0.333\linewidth]{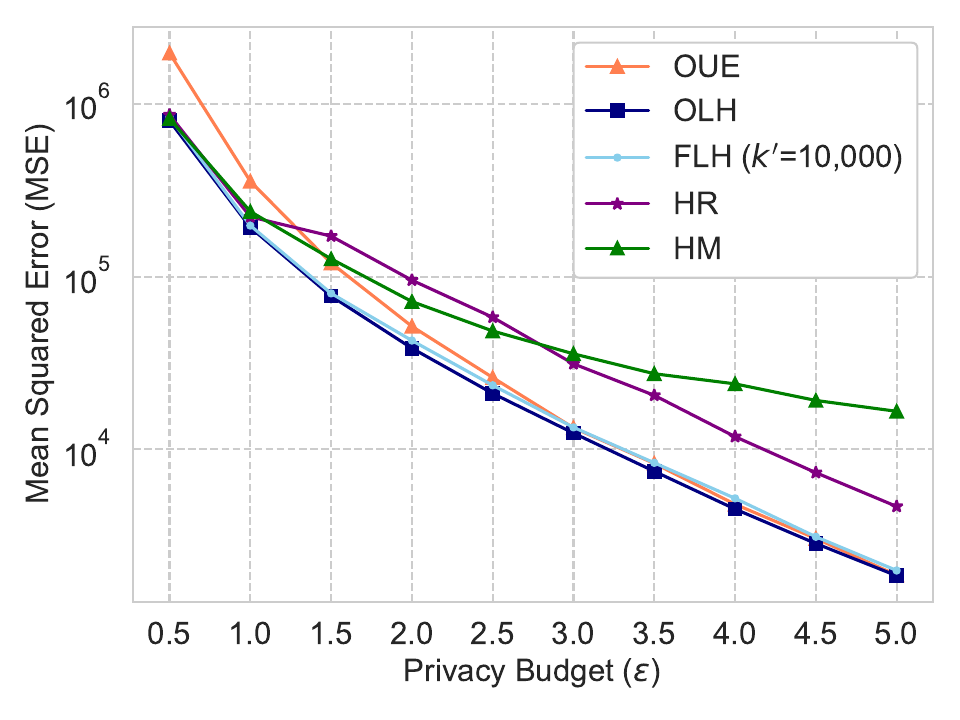}}
  \subfloat[\label{fig:group4_d} Varying $d$, fixing $\epsilon=3, n=100{,}000$]{%
        \includegraphics[width=0.333\linewidth]{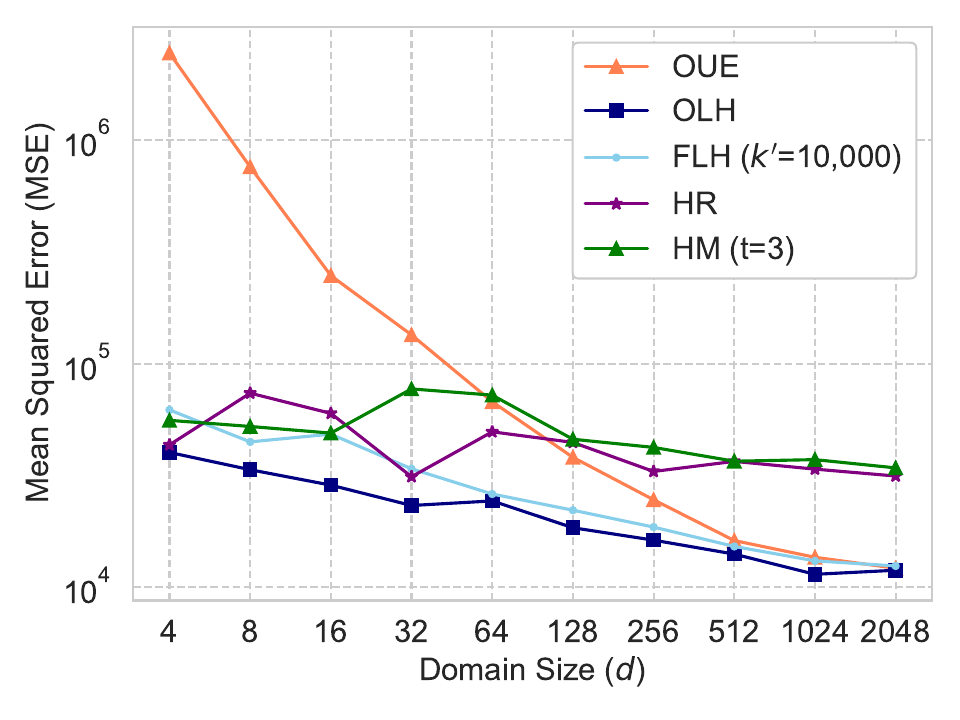}} 
  \subfloat[\label{fig:group4_time} Total time taken as $d$ varies]{%
        \includegraphics[width=0.333\linewidth]{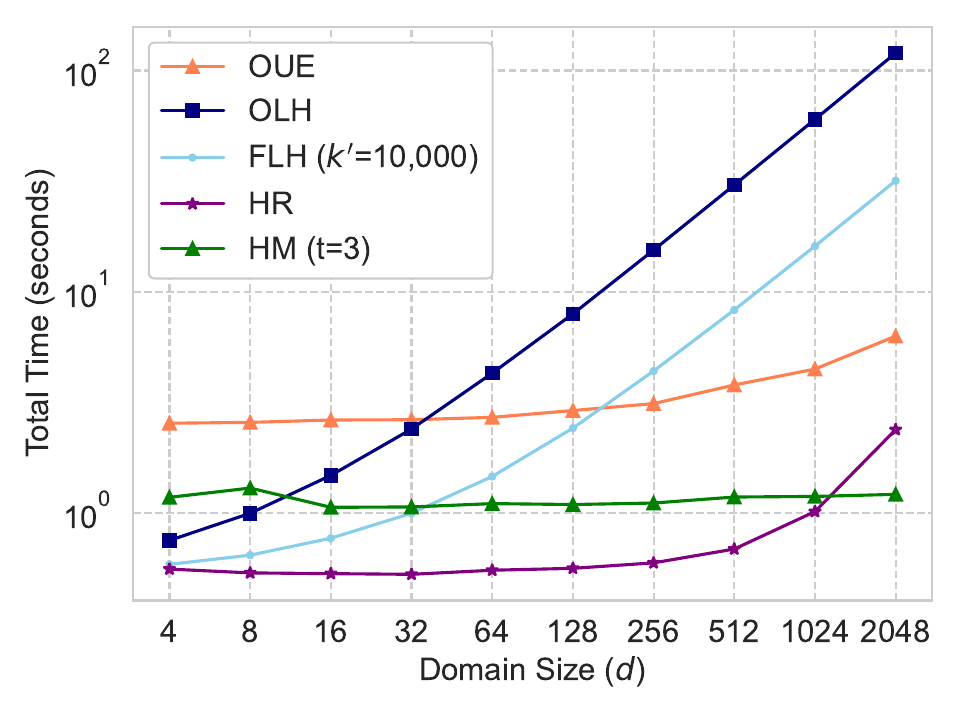}} 
  \caption{Comparison Experiments}
  \label{fig:group4} 
\end{figure*}

\subsection{Cross-comparison of Frequency Oracles}
\label{sec:expts:fo:comparisons}
In Figure \ref{fig:group4} we compare OUE, OLH, FLH, HR and HM, varying $\epsilon, d$  and looking at the total time taken. 
We omit DE since we have seen that its performance decreases  in larger domains, 
and BLH/SUE since OLH/OUE have better performance.  
For FLH we take $k^\prime=10{,}000$ and for HM we take $t=3$ when using $\epsilon=3$ otherwise we use the optimal $t$ values as found in Figure \ref{fig:group3_t}. 

In Figure \ref{fig:group4_eps}, we vary $\epsilon \in \{0.5, 1, \dots, 5\}$, fixing $d=1024, n=100{,}000$. 
We can see for low $\epsilon$ values (high privacy) that the techniques have comparable MSE values. 
As $\epsilon$ increases we can begin to see a split between the Hadamard technique (HR, HM) and the local hashing methods, with local hashing methods having lower MSE. 
Similarly in Figure \ref{fig:group4_d}, we vary $d \in \{2^{2}, \dots, 2^{11}\}$. We can see that as $d$ increases the split between the Hadamard and local hashing methods is more apparent. 
For very small domain sizes the techniques are comparable, but as $d$ grows large the local hashing methods have lower MSE.
Finally we look at the total time taken by the oracles as we vary $d$ in our experiments shown in Figure~\ref{fig:group4_time}. 
As noted previously, the local hashing methods are slow as $d$ increases and we can see the comparison in time between OLH and FLH with $k^\prime=10,000$. 
HR is the fastest technique, followed by HM and then OUE. 
We can clearly see that as the domain size increases, the local hashing methods are much slower. 

It should be noted that in these comparisons we use HM with $t=3$ where the aggregation time takes 3x longer than HM with $t=1$, since each user sends three Hadamard coefficients. For $t=1$, HM takes almost exactly the same total time as HR and there is little difference in time between them, except that HM's aggregation time increases as $t$ does. 
Out of all these techniques, OUE is the only one with  $O(d)$ communication cost and as $d$ grows further OUE would struggle to be competitive with the other techniques.

\begin{figure}[t]
  \includegraphics[width=0.85\linewidth]{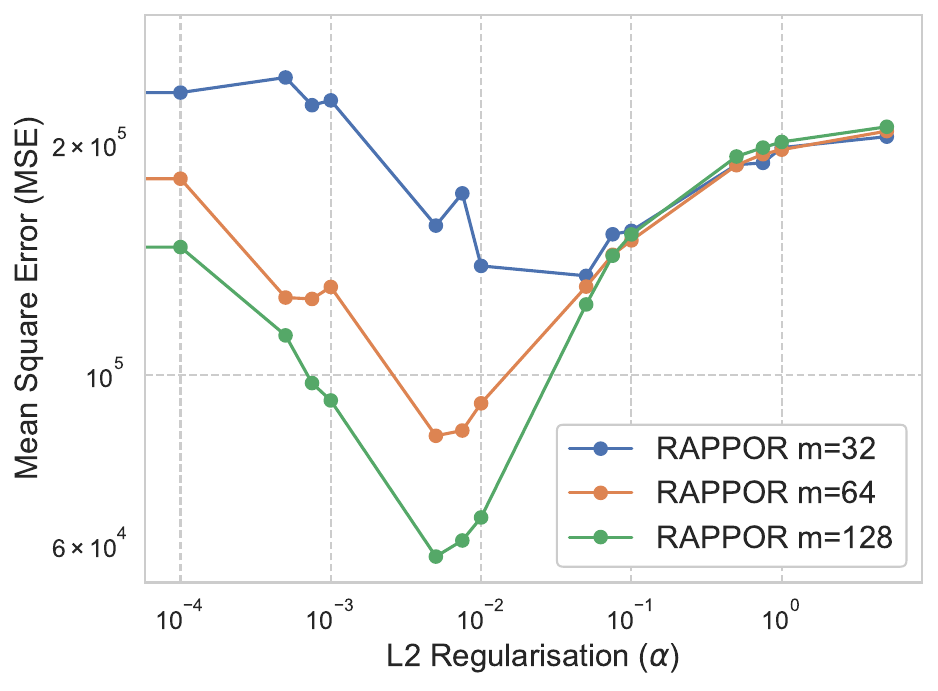}
  \caption{Varying the regularisation parameter $(\alpha)$ in RAPPOR, fixing $\epsilon =3, d=100{,}000, n=1{,}000{,}000$}
  \label{fig:group5_rappor} 
\end{figure}

\begin{figure*}[t]
 \subfloat[\label{fig:group5_sketch_m_mse} MSE varying $c$ fixing $r=32, \epsilon=3$]{%
      \includegraphics[width=0.3\linewidth]{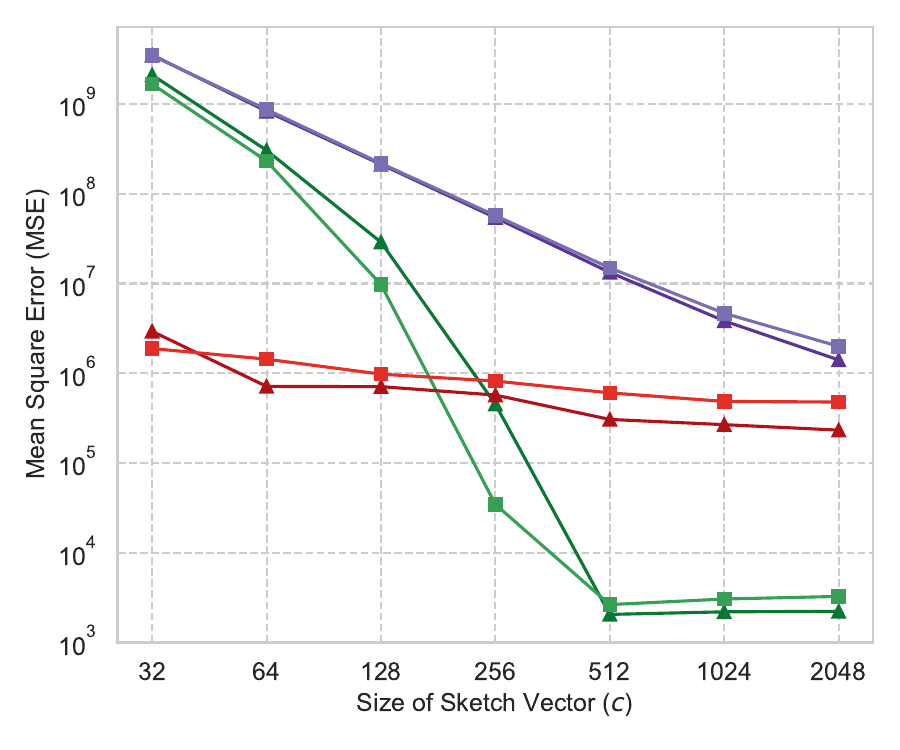}}
 \subfloat[\label{fig:group5_sketch_m_k_mse} $k$-MSE varying $c$ fixing $r=32, \epsilon=3$]{%
     \includegraphics[width=0.3\linewidth]{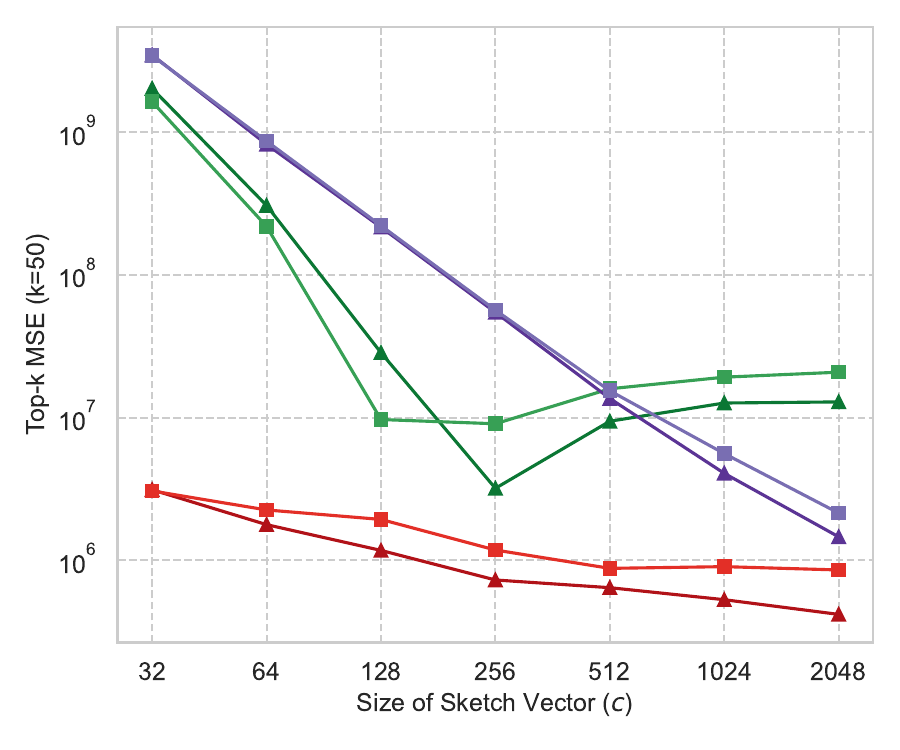}} 
  \subfloat[\label{fig:group5_sketch_k_mse} $k$-MSE varying $r$ fixing\newline $c=1024, \epsilon=3$]{%
        \includegraphics[width=0.4\linewidth]{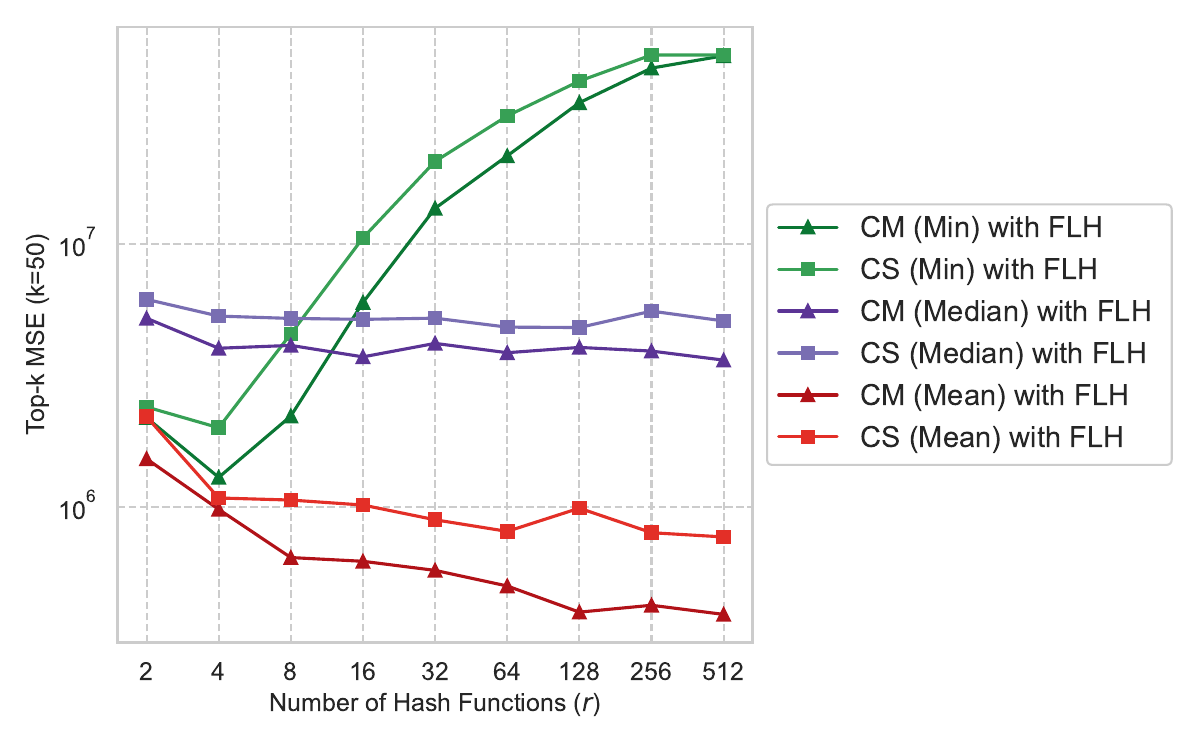}} 
  \caption{\edit{Varying Sketch Vector Size $(c)$ and Number of Hash functions $(r)$ with different sketch methods on the AOL dataset}}
  \label{fig:group5_sketch_params} 
\end{figure*}

\begin{figure*}[t]
  \includegraphics[width=0.85\linewidth]{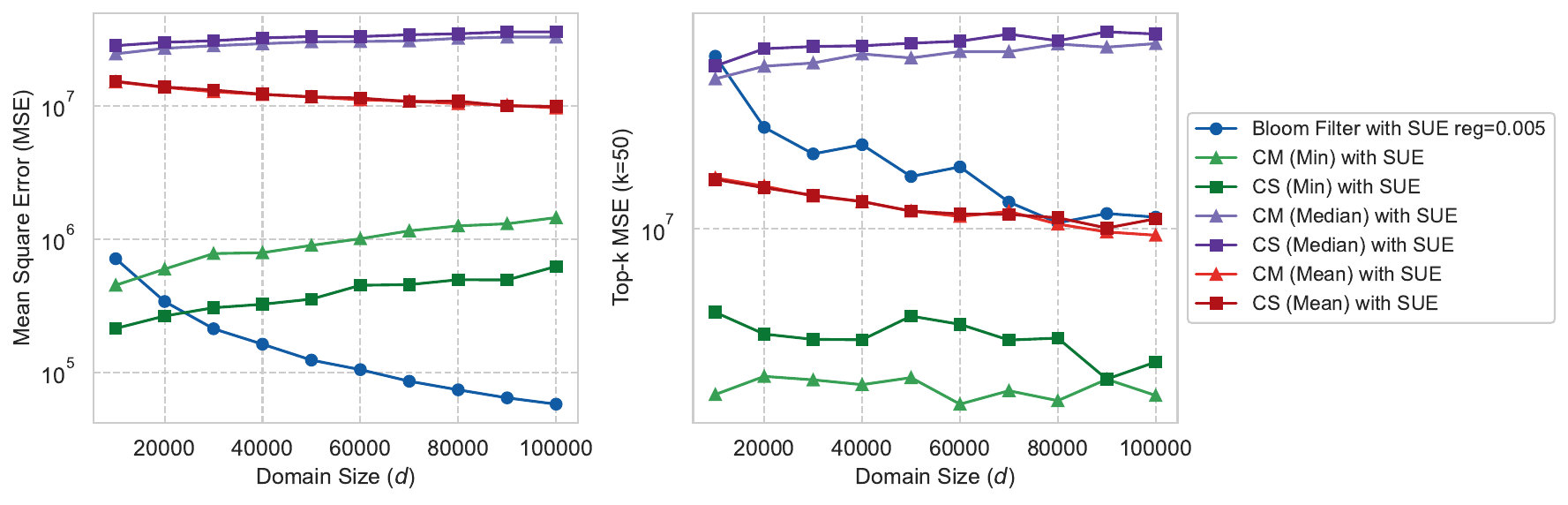} 
  \caption{MSE (left) and $k$-MSE (right) comparison of Bloom-filter, Count-Min Sketch (CM) and Count-Sketch (CS) with Symmetric Unary Encoding (SUE). Fixing $\epsilon =3,r=16, c=m=128$ \edit{on the Zipf distribution}}
  \label{fig:group5_bloom} 
\end{figure*}
\subsection{Domain Reduction Methods} 
\label{sec:expts:fo:sketches}
Next, we experimentally compare domain reduction methods (introduced in Section \ref{domain reduction}) which we combine with various FOs to allow us to efficiently estimate frequencies over very large domains. 
We devote particular attention to the following combinations:
\begin{itemize}
    \item Count-Sketch (CS) and Count-Min Sketch (CM) methods with FLH, Hadamard Response (HR), HadamardMech (HM)
    \item Bloom filters with Symmetric Unary Encoding (SUE) (the basis of the RAPPOR system~\cite{rappor})
    \item Apple's proposal of Count-Mean Sketch (CMS) as CM (Mean) sketch with OUE \footnote{Original implementations in \cite{appledp} actually utilise SUE. We use OUE which only involves changing the perturbation probabilities. In all other aspects the implementation details are identical to Apple’s CMS.} and Hadamard Count Mean Sketch (HCMS) as CM (Mean) with HM  ($t=1$)~\cite{appledp}
\end{itemize}
\edit{In these experiments, we utilise both synthetic Zipf data and real-world data based on a subset of AOL search queries in 2006\footnote{ \url{http://www.cim.mcgill.ca/~dudek/206/Logs/AOL-user-ct-collection/}}. 
The AOL dataset, among other things, contains both search queries by users and (if they clicked through) the URL of the website they chose. 
We focus only on the clicked URLs and not the search queries themselves. 
This forms a dataset of $n=1{,}935{,}614$ URLs, with $d=383{,}467$ unique URLs. 
For the Zipf data, we now increase the number of users to $n=1{,}000{,}000$ and vary $d$ from $10{,}000$ to $100{,}000$ in increments of $10{,}000$.}
Along with reporting MSE we also consider $k$-MSE which calculates the MSE over a subset of the top-$k$ most frequent elements in the domain. We fix $k=50$ in all experiments. We show this metric because as $d$ grows large, there will be many elements in the domain that have small (or close to $0$) frequency. Hence, if a frequency oracle was to underestimate frequencies (or predict close to $0$) it would have a (misleadingly) low MSE but a very high $k$-MSE as it would fail to accurately estimate the top-$k$ frequent elements. 
As we will see, there is often a trade-off in large domains between achieving a low MSE and a low $k$-MSE.

\subsubsection{Varying Parameters in RAPPOR}
The RAPPOR method combines Bloom filters with the Symmetric Unary Encoding (SUE) FO. 
One difficulty in using Bloom filters is that the estimation is not as straightforward as using sketches. 
The RAPPOR paper describes an estimation method using regularised regression for both selecting domain elements that could be frequent (via LASSO) and then estimating the frequencies of these elements (via non-negative ridge regression).

In Figure \ref{fig:group5_rappor}, we run an experiment varying the regularisation parameter $\alpha$ used in the ridge regression that estimates frequencies from the Bloom filters. 
We fix $n=1{,}000{,}000, \epsilon=3, d=100{,}000$ and plot three lines, each of which corresponds to a Bloom filter of size $m=32,64,128$. 
We can immediately note two things: Increasing the size of the Bloom filters results in lower MSE and that the correct choice of regularisation can lead to improved MSE performance.
The best regularisation value found was $0.005$ but values in the range $0.005-0.01$ have very similar MSE values. 
As we increase or decrease the regularisation from this point the MSE of RAPPOR begins to suffer.
The optimal value of regularisation is likely to be specific to the distribution of the data that is being collected, which may not be known in advance.

\subsubsection{Varying Sketch Parameters in Sketch Oracles}
One of the most popular deployments of sketch-based LDP frequency oracles are Apple's (H)CMS methods \cite{appledp}. The CMS technique can be thought of as a frequency oracle that combines a Count-Mean sketch with the OUE frequency oracle while its Hadamard variant is effectively the Hadamard Mechanism (HM) oracle combined with Count-Mean sketch. %
As explained in Section \ref{cm-sketch}, sketch methods consist of $r$ hash functions that map elements to a row of a $r \times c$ sized array. 
We therefore first look at the effect of varying these sketch parameters $r, c$. We consider both Count-Sketch (CS) and Count-Min sketch (CM) methods with variants that take either the mean, median or minimum of the sketch matrix to produce final frequency estimates. 

In Figure \ref{fig:group5_sketch_params} we vary the choice of sketch method \edit{on the AOL dataset with $\epsilon =3$}. We fix the frequency oracle to be FLH with $k^\prime=500$.
Figures~\ref{fig:group5_sketch_m_mse} and~\ref{fig:group5_sketch_m_k_mse} show both the MSE and $k$-MSE of the sketch methods while varying the size of the sketch vector $c \in \{32,64, \dots, 2048\}$ and fixing the number of hash functions to $r=32$. We can see that in general CM methods perform slightly better than CS methods, with a lower MSE. As we would expect, increasing the size of the sketch vector $c$ decreases the MSE across all methods. 
Figure~\ref{fig:group5_sketch_m_k_mse} shows an interesting trend in $k$-MSE for sketches that take the minimum estimate:
while fixing $r$ and increasing $c$, \edit{the $k$-MSE begins to slowly increase  after the $c=512$ point, at which the median methods begin to outperform the minimum}.

In Figure \ref{fig:group5_sketch_k_mse} we repeat the same experiment but vary the number of hash functions, $r \in \{2, 4, \dots, 512\}$ and fix $c=1024$. We only present the $k$-MSE graph for varying $r$, since we found that the MSE graph for varying $r$ showed very similar behaviour to Figure~\ref{fig:group5_sketch_m_mse} with a general decrease in MSE as the parameter is increased. 
Broadly, we see a similar pattern to the previous experiment---with CM methods slightly outperforming the CS methods. However, the notable difference is in the $k$-MSE graph, where increasing $r$ results in worse $k$-MSE performance for sketch methods that take the minimum as their estimate. 

\begin{figure*}[t]
    \includegraphics[width=0.8\linewidth]{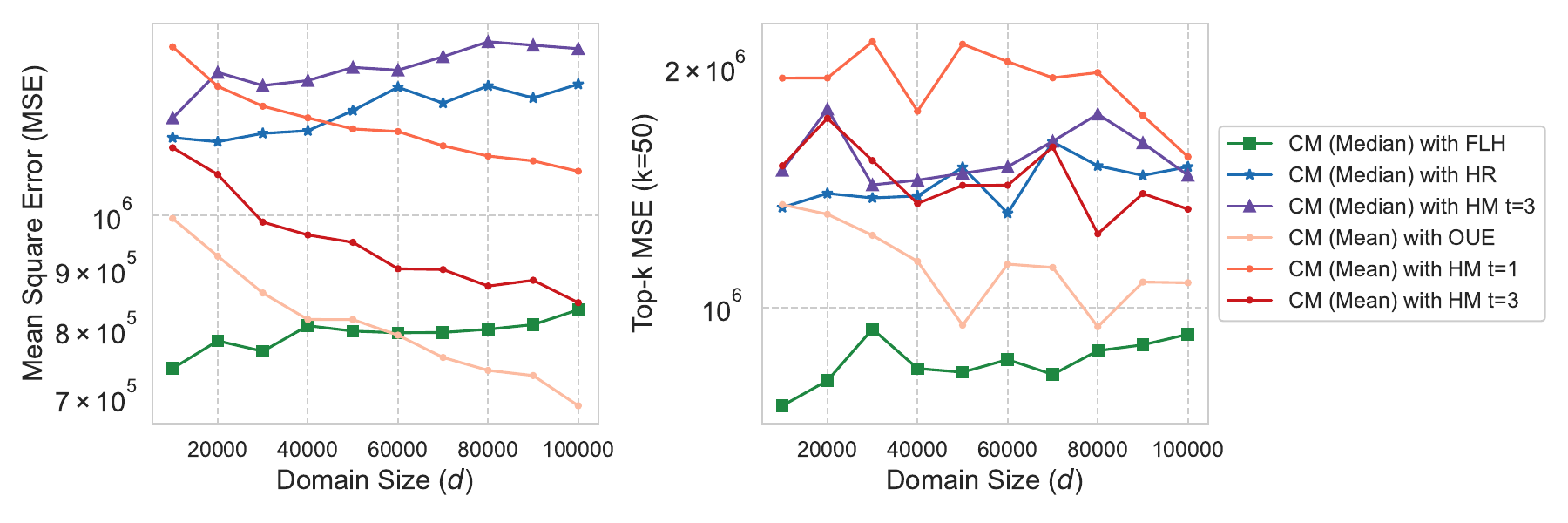}
    \caption{MSE (left) and k-MSE (right) comparing Count-Median sketch methods (with FLH, HR, HM) and Apple's CMS/HCMS while varying $d$, fixing $\epsilon=3, n=1{,}000{,}000, r=32, c=1024$ \edit{on the Zipf distribution}}
  \label{fig:group5_sketch_oracles} 
\end{figure*} 

\subsubsection{Comparing Sketch Estimation Methods}
\label{sketch-estimates}
Across all of Figure~\ref{fig:group5_sketch_params} we can see that the choice of sketch estimate quite clearly affects both the MSE and $k$-MSE performance. 
We see that using sketches that take the Minimum as the estimate have the lowest MSE followed by Mean and then the Median. 
However, the picture becomes more complex when we look at the $k$-MSE plot in Figure~\ref{fig:group5_sketch_k_mse}. 
Here the results are somewhat reversed, with Min methods performing the worst \edit{(when $c$ is large)} followed by Median and then Mean. As mentioned, as the number of hash functions $r$ is increasing, the $k$-MSE of the Minimum methods begin to increase.

The likely reason for this is that as the number of hash function increase, the number of users are effectively split into cohorts where their data contributes to a single row of the sketch matrix. 
This means as $r$ increases, the user base is split between $r$ frequency oracles each of which estimate a single row in the sketch. This will lead to noisy estimates that underestimate the true frequencies and inflate MSE. 
Generally, it seems that for small $r$ and large $c$ the Mean and Median are comparable and achieve the best balance between MSE and $k$-MSE.

\subsubsection{Sketching vs Bloom Filters}
In Figure~\ref{fig:group5_bloom} we compare the various sketching methods with the Bloom filter approach (implemented via RAPPOR). Since RAPPOR can be thought of as a Bloom filter combined with the Symmetric Unary Encoding (SUE) frequency oracle, we also compare sketching methods that use SUE as their oracle. To ensure a fair comparison we fix the sketch sizes to $r=16, c=128$ to match that of a Bloom filter with size $m=128$. For the Bloom filter we also use $k=2$ hash functions and $8$ cohorts. We fix the regularisation parameter to be $\alpha=0.005$ as informed by our previous experiments. 
We fix $\epsilon =3, n=1{,}000{,}000$ and vary $d$ \edit{on Zipf distributions}. We observe generally that in this small sketch regime the Bloom filter approach is certainly competitive. It achieves MSE that is similar CS (Min) and begins to outperform it as $d$ increases. %

Interestingly, when we look at this small sketch regime taking the minimum as the sketch estimate leads to better performing oracles in both MSE and $k$-MSE, different to what we saw for much larger sketches. 
Since we have a very large domain and a small sketch, there is likely to be many collisions in the sketch. 
Taking the minimum seems to be more robust to collisions in the sketch whereas the mean and median struggle with this on smaller sketches.
There are limitations to the Bloom filter approach. %
The difficult choice of regularisation and the complicated estimation procedure makes implementation fairly unattractive and with larger sketches CM/CS can achieve equal MSE to the Bloom filter approach.
Consequently, we drop the Bloom filter approach from subsequent experiments.

\begin{figure}[t]
    \centering
    \includegraphics[width=0.85\linewidth]{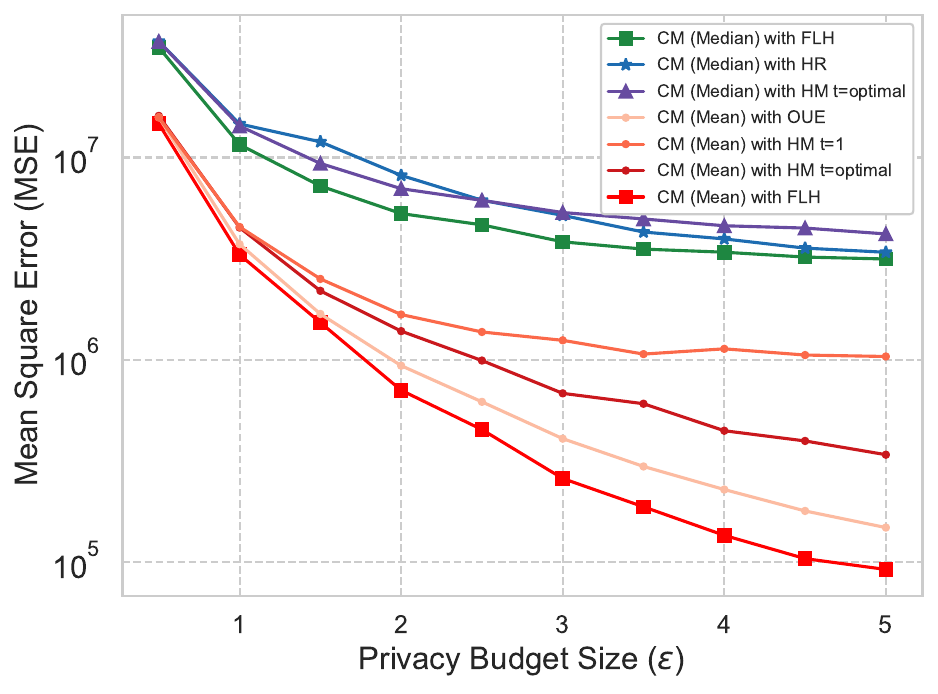} %
    \caption{\edit{Varying $\epsilon$ on the AOL dataset ($d=330,000$) for Count-Median and Mean Sketch Methods ($r=32, c=1024$)}}
     \label{fig:group5_fo_eps} 
\end{figure}

\begin{figure*}[t]
  \includegraphics[width=0.8\linewidth]{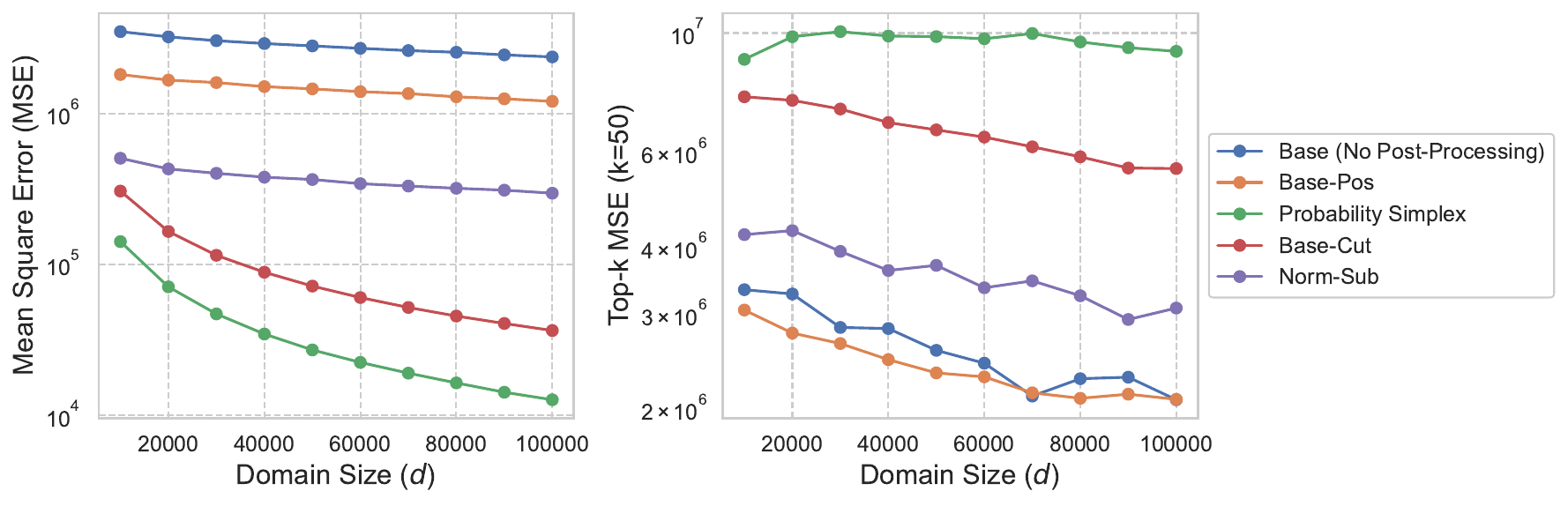}
  \caption{MSE (left) and $k$-MSE (right) of various post-processing techniques on FLH, varying $d$ and fixing $\epsilon=3, n=1,000,000$}
  \label{fig:group6} 
\end{figure*}

\subsubsection{Comparing Sketching Using Different Frequency Oracles}
We now compare the sketching methods while varying the frequency oracles used to privatise data. Since our previous experiments have shown that Count-Sketch methods perform slightly worse, we omit them from our comparison. 
Instead we compare Apple's CMS and HCMS (as CM (Mean) OUE and CMS (Mean) HM ($t=1$), respectively) and CM (Median) with FLH, HR and HM as its oracle. For HM we fix $t=3$ when $\epsilon=3$ otherwise we use optimal $t$ values found from our previous experiments. We also plot CM (Mean) HM ($t=3$) to see if there is direct improvement over Apple's HCMS.
The experimental results are shown in Figure \ref{fig:group5_sketch_oracles} where we fix $\epsilon=3, r=32, c=1024, n=1{,}000{,}000$ and vary the domain size.

We can see that CM (Mean) OUE and CM (Median) FLH achieve the lowest MSE and when looking at the $k$-MSE it seems that in general the CM (Median) methods have lower $k$-MSE than CM (Mean). We also note that CM (Mean) with HM $(t=3)$ outperforms taking CM (Mean) HM $(t=1)$ and therefore improvements in Apple's current deployment of HCMS could be made by modifying their protocol to  sample more Hadamard coefficients.
Additionally, the CM sketching oracles have a logarithmic communication cost of $O(\log c)$ which for large sketches is much smaller than CM (Mean) OUE's $O(c)$ communication. It seems that the CM (Median) FLH method has similar performance to CM OUE, yet achieves smaller communication cost without loss of accuracy. 
By expressing Apple's (H)CMS protocols in our framework we have highlighted some modifications and variations that lead to increased performance.

\edit{\subsubsection{Comparing Sketching on the AOL Dataset} \label{aolfo}
As a final test for the sketching oracles, we  experiment on the real-world AOL search query dataset. 
We use the same sketching oracles as in Figure \ref{fig:group5_sketch_oracles}, but additionally include CM (Mean) FLH, all with $\epsilon=3$ (Figure \ref{fig:group5_fo_eps}). 
We observe that CM (Mean) FLH has the lowest MSE followed by (Mean) OUE and then (Mean) HM methods. Both Mean and Median methods are grouped closely together in performance when $\epsilon$ is small, but as it increases the MSE gap between protocols becomes clearer. 
Note that there is a notable gap between Mean and Median methods on this dataset. 
These findings are consistent with the message of Figure~\ref{fig:group5_bloom}, that we can achieve good performance (on both synthetic and real-world data) with techniques like CM (Mean) FLH and HM that achieve smaller communication cost than current state-of-the art methods (namely Apple's CMS).}

As we have seen, when we increase $k^\prime$ in the FLH oracle, the MSE is reduced. We chose $k^\prime=500$ to ensure that sketching with FLH, HR, HM and OUE take roughly the same amount of total time. 
Increasing $k^\prime$ makes the overall oracle slower but can increase accuracy further. 
In general, all sketching methods we have compared have almost identical client and server-side processing times.

\subsection{Post-Processing Experiments}
\label{sec:expts:post}
We next run experiments using the post-processing methods described in Section~\ref{post-processing}. 
Figure \ref{fig:group6} shows the results, fixing $d=100{,}000, \epsilon=3, n=1{,}000{,}000$ and varying the post-processing methods used on a FLH oracle with $k^\prime=500$.

Just like sketching, the different post-processing methods also have a MSE/$k$-MSE trade-off. 
As we would intuitively expect, 
\edit{Base-Pos} results in better MSE and $k$-MSE performance than performing no post-processing at all. 
\edit{Norm-Sub and Base-Cut} strike a balance between MSE and $k$-MSE. 
\edit{For Norm-Sub} it seems that the combination of rounding negative values and normalising individual estimates to match the total number of users helps to decrease MSE while still retaining accurate top-$k$ estimates ensuring a relatively low $k$-MSE. 
For \edit{Base-Cut}, we achieve a lower MSE by rounding values after the threshold down to $0$ but during this process we seem to round down certain top-$k$ values which inflates the $k$-MSE.
For large domains projecting the results onto a probability simplex pushes the estimates down, resulting in a low MSE but a very high $k$-MSE since frequent domain items have their estimates reduced. 

Overall it seems that \edit{Norm-Sub} has the best balance between MSE/$k$-MSE, but \edit{Base-Pos} also performs well and is clearly a better choice than no post-processing. \edit{These conclusions are in line with the experimental observations of \cite{wangConsistency}.}
Note that for sketching oracles, we can consider two areas to perform post-processing, one that can be performed on the internal frequency oracles that estimate counts in sketch rows, and a second post-processing step over the entire sketch oracle itself when estimating the domain. 
Our tests showed that taking combinations of \edit{Norm-Sub} in both of these give the best balance of $k$-MSE/MSE, but we omit these plots here.

\begin{table*}[t]
\centering
\footnotesize
\caption{Top 20 HH + FO Combinations on the AOL dataset based on F1 scores. The standard deviations of F1 scores are presented in parentheses. The maximum values of precision, recall and F1 are in bold. \label{tab:HH}}

\begin{tabular}{lllllrrl}
\toprule
{} & Heavy Hitter & Sketch Method & Frequency Oracle &            Parameters &  Precision &  Recall &       F1 Score \\
\midrule
1  &          PEM &   CM (Median) &              FLH &  T=20, r=1024, c=2048 &      0.822 &    0.74 &  \textbf{0.779} (0.052) \\
2  &          PEM &   CM (Median) &              FLH &   T=20, r=128, c=1024 &      0.789 &    0.74 &  0.763 (0.051) \\
3  &          PEM &     CM (Mean) &              OUE &  T=10, r=1024, c=2048 &      0.626 &    \textbf{0.96} &  0.757 (0.057) \\
4  &          PEM &     CM (Mean) &              OUE &    T=20, r=32, c=1024 &      0.782 &    0.72 &  0.749 (0.025) \\
5  &          PEM &   CM (Median) &              FLH &  T=10, r=1024, c=2048 &      0.758 &    0.74 &  0.748 (0.061) \\
6  &          PEM &     CM (Mean) &              OUE &   T=20, r=128, c=1024 &      0.786 &    0.70 &  0.738 (0.035) \\
7  &          PEM &     CM (Mean) &              OUE &    T=10, r=32, c=1024 &      0.782 &    0.70 &  0.738 (0.025) \\
8  &          PEM &   CM (Median) &              FLH &    T=20, r=32, c=1024 &      0.792 &    0.68 &  0.731 (0.036) \\
9  &          PEM &     CM (Mean) &              OUE &  T=20, r=1024, c=2048 &      0.766 &    0.70 &   0.73 (0.029) \\
10 &          PEM &   CM (Median) &              FLH &    T=10, r=32, c=1024 &      0.744 &    0.70 &  0.721 (0.055) \\
11 &          PEM &   CM (Median) &              FLH &   T=10, r=128, c=1024 &      0.552 &    0.88 &  0.677 (0.031) \\
12 &          PEM &     CM (Mean) &              OUE &   T=10, r=128, c=1024 &      0.541 &    0.88 &  0.669 (0.076) \\
13 &          PEM &     CM (Mean) &         HM (t=1) &   T=20, r=128, c=1024 &      0.616 &    0.60 &  0.607 (0.098) \\
14 &          PEM &     CM (Mean) &         HM (t=1) &  T=20, r=1024, c=2048 &      0.580 &    0.56 &  0.569 (0.093) \\
15 &          PEM &     CM (Mean) &         HM (t=1) &    T=10, r=32, c=1024 &      0.551 &    0.54 &    0.545 (0.1) \\
16 &          PEM &     CM (Mean) &         HM (t=1) &    T=20, r=32, c=1024 &      0.516 &    0.48 &  0.496 (0.052) \\
17 &          SFP &     CM (Mean) &              OUE &   T=10, r=128, c=1024 &      0.967 &    0.32 &   0.465 (0.12) \\
18 &          SFP &     CM (Mean) &              OUE &  T=10, r=1024, c=2048 &      0.960 &    0.30 &   0.45 (0.065) \\
19 &          SFP &     CM (Mean) &              OUE &  T=20, r=1024, c=2048 &      \textbf{1.000} &    0.28 &  0.436 (0.051) \\
20 &          PEM &     CM (Mean) &         HM (t=1) &   T=10, r=128, c=1024 &      0.322 &    0.62 &  0.424 (0.046) \\
\edit{$\cdots$} & \edit{$\cdots$} & \edit{$\cdots$} & \edit{$\cdots$} & \edit{$\cdots$} & \edit{$\cdots$} & \edit{$\cdots$} & \\
\edit{37} & \edit{TH} & \edit{CM (Median)} & \edit{FLH} & \edit{T=10, r=128, c=1024}  & \edit{0.202} & \edit{0.18} & \edit{0.189 (0.068)} \\
\edit{$\cdots$} & \edit{$\cdots$} & \edit{$\cdots$} & \edit{$\cdots$} & \edit{$\cdots$} & \edit{$\cdots$} & \edit{$\cdots$} & \\
\bottomrule
\end{tabular}

\end{table*}
\section{Heavy Hitter Experiments}
\label{sec:hhexpts}

We conclude our experiments by applying the frequency oracles to the heavy hitter (HH) problem. We consider three popular implementations of LDP HH techniques, Apple's Sequence Fragment Puzzle (SFP) \cite{appledp}, and two from the Hierarchical Search framework: Prefix Extending Method (PEM) \cite{pem} and TreeHistogram (TH)~\cite{practicalhh,bassilysmith}. 

\para{Heavy Hitters Dataset}
We base our HH experiments on a subset of the 2006 AOL search query dataset, \edit{as introduced in Section \ref{aolfo}. Again, we focus only on the clicked URLs and not the search queries themselves }. This forms a dataset of $1{,}935{,}614$ URLs, with $383{,}467$ unique URLs. The goal of applying HH techniques in this scenario is to discover the most popular URLs that were clicked by users, examples of which are websites like \say{google.com}, \say{yahoo.com} and \say{myspace.com}. 
Additional pre-processing was performed to strip website URLs of leading domain prefixes like \say{www.} and \say{https://} but domain suffixes like \say{.com} were left untouched.

\subsection{Experiment Parameters}
We run combinations of heavy hitter search protocols (PEM, SFP, TH) with various sketch-based frequency oracles with the goal of finding the best combination of frequency oracle and heavy-hitter search technique for this dataset. We run the HH + FO combinations to attempt to find the top-10 most frequent URLs in the dataset. 
We privatise all $1{,}935{,}614$ URLs with a privacy budget of $\epsilon=3$, which for SFP and TH is split evenly into $\epsilon=1.5$ for both the fragment and word estimator. We measure the performance via the precision, recall and the associated F1 score calculated by comparing the strings found to the true top-10 frequent URLs in the dataset. 

We truncate all URLs in the dataset to a length of $6$ characters. Across all the HH protocols, we fix the length of prefixes that are used to build up the HHs to be a length of $2$ characters. 
We run two variations of the HH protocols for $T=10, 20$ where $T$ denotes the number of top-T prefixes that are used to build up the heavy hitters at each stage of the protocol. 
While each of the three HH protocols differ in how they do this, all of them estimate some combination of top-$T$ prefixes from which they build up their HH predictions. For the frequency oracles we choose Apple's CMS (as CM (Mean) + OUE), HCMS (as CM (Mean) + HM) and CM FLH (Median) to compare with current HH deployments that use (H)CMS. %
We use three sketch-size variants  $(r,c)$: $(32,1024), (128,1024), (1024,2048)$. Overall, this results in a total of $54$ combinations, each of which is repeated $5$ times with the results averaged. 

\subsection{Results}
\label{sec:expts:hh:results} 
In Table \ref{tab:HH}, we present the top-20 combinations of HH + FO sorted by their F1 score. \edit{Since no instantiation of TH appears in the top-20, we have also presented the first occurrence of TH in the table.} The most striking result is that PEM dominates SFP and TH in terms of the F1 score, no matter the oracle that we choose. These results corroborate similar experiments in \cite{pem}, however they choose to just fix the oracle to be OLH and compare PEM with slight variations of hierarchical search techniques. 

Focusing on the PEM results, CM FLH has the best F1 performance for the two larger sketch-size variations with $T=20$, followed by CMS (i.e., CM (Mean) with OUE) with the largest sketch-size variant. 
While the F1 scores of using CMS and CM FLH with PEM on this dataset are very comparable, the most notable result here is that we can achieve slightly better F1 scores than CMS with both much smaller sketch-sizes and much lower communication cost by using CM FLH. To achieve an equivalent communication cost we would have to use HCMS which as we can observe performs notably worse than CM FLH. 

The table suggests that combining SFP \edit{+} CMS outperforms SFP \edit{+} CM FLH. 
Since SFP splits the privacy budget between estimating fragments and words, then both oracles in our experiments use $\epsilon=1.5$. 
Our experiments in Figure \ref{fig:group5_fo_eps} show that for lower privacy budgets, CMS outperforms \edit{(Median)} FLH and this is the likely cause to the better performance of SFP + CMS. 
Just as is the case for PEM, we see that CM FLH outperforms HCMS when combined with SFP as well.

\section{Concluding Remarks}
\label{sec:concs}

We have seen that methods for collecting frequency statistics under the model of local differential privacy can be practical and accurate.
Although there are many variations, by viewing them through our four layer model, it is possible to navigate the large space of possibilities to identify the most effective combinations. 
Our experimental findings can be summarized as follows: 

\para{Frequency oracles}
For small domains, up to hundreds or thousands in size, methods based on local hashing are most effective. 
Our proposed ``fast'' local hashing (FLH) approach is several times faster than optimal local hashing (OLH), without much loss of accuracy. 
For larger domains, Hadamard-based approaches, particularly Hadamard response (HR) is almost as accurate but can be orders of magnitude faster. 

\para{Domain size reduction}
For larger domains, domain size reduction via sketches works well. 
Count-min sketch based approaches seem marginally preferable to Count sketch ones, using estimation with median or mean performing best across a range of domain sizes up to hundreds of thousands. 

\para{Post-processing}
Post-processing reduces error, with the method of rounding negative values to zero being simplest, while adding a small normalization constant proving a good choice when feasible. 

\para{Heavy hitters}
Combining all these to solve heavy hitters problems, the combination of sketching with local hashing seems the best overall, though other similar parameter settings do almost as well.  
The hierarchical encoding, in particular the prefix encoding method (PEM) clearly dominates other choices. 

\smallskip
Last, while we focus on frequency statistics, it is natural to extend these methods to related questions (e.g., quantiles and range queries), using transformations that are now standard~\cite{rangequeries}. 

\para{Acknowledgments}
This work was supported by the UK government and European Research Council grant ERC-2014-CoG 647557.

\bibliographystyle{ACM-Reference-Format}
\bibliography{ldphh}

\end{document}